\documentclass[12 pt]{article}
 \usepackage{amsmath, amsthm, amssymb, bigints, bm, setspace}
 \usepackage{color}
 \usepackage[margin=1.0in]{geometry}
  
\newcommand{\bfb} {\mbox{\boldmath ${\bm \beta}$}}

\newcommand{\bfl} {\mbox{\boldmath $\lambda$}}

\newcommand{\bfe} {\mbox{\boldmath $\epsilon$}}

\newif\ifpdf
	\ifx\pdfoutput\undefined
	\pdffalse 
	\else
	\pdfoutput=1 
	\pdftrue
	\fi

	\ifpdf
	\usepackage[pdftex]{graphicx}
	\else
	\usepackage{graphicx}
	\fi

\begin{document}

\ifpdf
	\DeclareGraphicsExtensions{.pdf, .png, .jpg, .tif}
	\else
	\DeclareGraphicsExtensions{.eps, .jpg}
	\fi

\begin{center}
{\Large {\bf Bayes Regularized Graphical Model Estimation in High Dimensions\\}} 
Suprateek Kundu, Veera Baladanyuthapani, and Bani K. Mallick. \\
\end{center}
{\noindent {\bf Abstract:}} There has been an intense development of Bayes graphical model estimation approaches over the past decade - however, most of the existing methods are restricted to moderate dimensions. We propose a novel approach suitable for high dimensional settings, by decoupling model fitting and covariance selection. First, a full model based on a complete graph is fit under novel class of continuous shrinkage priors on the precision matrix elements, which induces shrinkage under an equivalence with Cholesky-based regularization while enabling conjugate updates of entire precision matrices. Subsequently, we propose a post-fitting graphical model estimation step which proceeds using penalized joint credible regions to perform neighborhood selection sequentially for each node. The posterior computation proceeds using straightforward fully Gibbs sampling, and the approach is scalable to high dimensions. The proposed approach is shown to be asymptotically consistent in estimating the graph structure for fixed $p$ when the truth is a Gaussian graphical model. Simulations show that our approach compares favorably with Bayesian competitors both in terms of graphical model estimation and computational efficiency. We apply our methods to high dimensional gene expression and microRNA datasets in cancer genomics. 

\newpage

\begin{center}{\bf 1. INTRODUCTION }  \end{center} 

Recent technological advances in many scientific disciplines such as genomics, imaging and environmental studies result in data sets with very many variables. A convenient framework for analyzing and interpreting relationships between the variables is via graphical models. Graphical models are useful tools for detecting a network of dependencies amongst a group of $p$ measurements denoted by $x_1,\ldots,x_p$. In particular under a Gaussian set-up, the pattern of zeros in the inverse covariance or the precision matrix correspond to conditional dependency interpretations between $x_1,\ldots,x_p$. Covariance selection (Dempster, 1972) refers to the approach of estimating these structural zeros from the data. Our focus in this article is to propose a novel, flexible and scalable Bayesian covariance selection strategy in high dimensional fixed $p$ settings (our examples include $p$ in several hundreds).

In high-dimensional settings, traditional methods for covariance/ precision matrix estimation  (and hence graphical model estimation) may not perform well (Whittaker 1990; Edward 2000), prompting the development of a number of frequentist approaches (Meinshausen and Buhlmann, 2006; Yuan and Lin; 2007; Peng et al., 2009). 
From a Bayesian perspective, there has been an intense methodological development of covariance selection approaches, in particular on prior constructions for graphical models based on discrete mixture formulations. Let ${\bf X} = (X_1,\ldots,X_n)^T = (X_{c1},\ldots,X_{cp})$ be the $n\times p$ dimensional data matrix, with subscript $c$ denoting the columns.  Usual Bayesian covariance selection approaches  focus on  the following discrete mixture formulation:
\begin{eqnarray}
X_l \sim N(\theta,\Sigma_G), \quad \Sigma_G\sim\pi(\Sigma|G), \quad \theta \sim \pi(\theta), \quad G\sim \pi(G), \quad l=1,\ldots,n, \label{eq:base1}
\end{eqnarray}
where the graph $G$ is defined using a set of nodes or vertices $V=\left\{1,\ldots,p\right\}$ and an edge set $E=(e_{ij})$ with $e_{ij}=1$ if and only if the $(i,j)$th entry of the precision matrix is non-zero. For a fixed graph $G$, the support of $\Sigma^{-1}_G$ is the cone $M^+_G$, the space of all positive definite matrices having exact zeros for off-diagonals corresponding to absent edges. Here $\theta$ denotes the mean which is usually set to zero after standardizing the measurements. 

Typical prior choices include discrete mixture priors such as the hyper inverse Wishart prior (Dawid and Lauritzen, 1993) for the covariance or the $G$-Wishart prior for the precision (Diaconnis and Ylvisaker, 1979; Roverato, 2000; Atay-Kayis and Massam 2005). The implementation of most of these approaches rely on reversible jump Markov chain Monte Carlo (MCMC) algorithms (Giudici and Green, 1999; Dellaportas, Giudici, and Roberts, 2003; Wong, Carter, and Kohn, 2003). These algorithms explore the graph space and subsequently select graphs with high posterior probabilities $P(G|{\bf X})$ or estimate a graph by including edges having posterior inclusion probability $>0.5$. Jones et al. (2005) proposed the shotgun stochastic search algorithm designed to efficiently move toward regions of high posterior probability in the model space using a parallel computing approach, while Scott and Carvalho (2008) developed a greedy approach called the feature inclusion search (FINCS) algorithm for decomposable Gaussian graphical models. As an alternative to discrete mixture priors, Wang (2012) recently proposed the Bayesian graphical lasso which discovers absent edges by shrinking the corresponding off-diagonals towards zero, and subsequently uses a heuristic post-MCMC processing step to estimate the edge set. 

As $p$ increases, the cardinality of the graph space increases exponentially, making it computationally intractable if not impossible for many MCMC based approaches to efficiently explore the graph space. This problem is somewhat akin to known difficulties encountered by stochastic search variable selection (George and McCulloch, 1993) approaches to navigate the model space for high dimensional regression settings (Bondell et al., 2012; Kundu et al., 2013). However, the problem is far more severe under reversible jump MCMC approaches for graphical model estimation, as the graph space (having cardinality $2^{p(p-1)/2}$) can become huge even for moderate $p$. As a result usual discrete mixture based approaches can fail to discover models with high posterior probabilities, with the edge-specific posterior inclusion probability estimates being potentially unstable under finite runs of the MCMC chain (as demonstrated in our simulations). Moreover in high dimensions, the results can be sensitive to the choice of the prior on the graph space. 

In addition, for non-decomposable graphs, estimation of graph posterior probabilities require computing $P(X|G)$ by marginalizing out the nuisance parameters, which entails computationally involved heuristic approximations (Atay-Kayis and Massam, 2005; Lenkoski and Dobra, 2011), and makes application to higher dimensions increasingly difficult, if not impossible. Due to such computational considerations, discrete mixture based approaches often have to restrict their attention to the class of decomposable graphs. On the other hand Bayesian graphical lasso (Wang, 2012) entails severe computational burden for increasing $p$, due to column-wise updates required for sampling the precision matrix (as we evidence in the simulation section).


To address these issues, we propose a novel approach for graphical model estimation in Gaussian graphical models which is suitable for high dimensional settings. The proposed approach separates model fitting and covariance selection. First, the full model based on a complete graph is fit under a class of novel continuous shrinkage priors which is denoted as Regularized Inverse Wishart (RIW) priors in the sequel, and is based on mixtures of inverse Wishart priors on the precision matrix. Our approach is novel in assigning suitable priors on the scale parameters, marginalizing over which induces adaptive shrinkage on the precision matrix elements. The shrinkage is induced under equivalence with known Cholesky-based regularization (Pourahmadi, 1999; Chen and Dunson, 2003; Wu \& Pourahmadi, 2003) employing a group lasso (Yuan and Lin, 2006) type penalty. However unlike usual Cholesky-based regularization approaches, the proposed prior is order invariant and enables conjugate updates of the precision matrix, leading to efficient posterior computation. Due to order invariance, the proposed approach can be applied to a broad spectrum of problems, and is not restricted to scenarios where there is a natural ordering among variables.  

Subsequently we propose a post-MCMC model selection step, which uses $L_0$ penalized joint credible regions to perform neighborhood selection for each node, thus resulting in graphical model estimation. The merits of $L_0$ penalized credible regions in the variable selection context have been discussed by many authors (Schwarz, 1978; Liu and Wu, 2007; Kim et al., 2012; Shen, Pan and Zhu, 2012; Bondell et al., 2012), but to our knowledge we are the first to apply it in graphical model estimation context. The penalized approach can be implemented efficiently using existing algorithms such as LARS (Efron et al., 2004) and has attractive theoretical justifications in terms of graphical model selection consistency in recovering the true Gaussian graphical model, under suitable assumptions. While selection consistency is highly desirable for graphical model estimation approaches, to our knowledge such results are sparse in the Bayesian paradigm, a gap we bridge in this work.


In summary, the proposed approach overcomes several difficulties associated with existing Bayesian alternatives: (1) it obviates having to specify prior probabilities on the graph space which can significantly influence the final results under discrete mixture alternatives, especially for large $p$, (2) it does not require long runs of MCMC chains to search over model space and the posterior computation proceeds using a straightforward fully Gibbs sampler, (3) it can sample entire precision matrices as a whole due to conjugacy, thus attaining substantial computational gains and being scalable to high dimensions, and (4) it is applicable to a broad class of models including both decomposable and non-decomposable graphs. 

The paper is structured as this. In Section 2 we propose the RIW prior formulation and explore its properties including connections to Cholesky-based regularization approaches, as well as describe the associated posterior computation. In Section 3, we describe the model/covariance selection approach, and establish selection consistency of the proposed approach. In Section 4, we describe our numerical simulation studies and Section 5 illustrates an application in gene/microRNA regulatory networks in cancer. Section 6 includes additional discussions and all technical details are collected into an Appendix. \\

\vskip 12pt

\begin{center} {\bf 2. SHRINKAGE PRIORS FOR PRECISION MATRICES} \end{center} 

{\noindent {\bf 2.1. The Regularized Inverse Wishart prior}}

In this section, we propose shrinkage priors on the precision matrix, which are based on mixtures of inverse Wishart formulation on the covariance. Without loss of generality we assume a zero mean model, i.e. set $\theta=0$ in (\ref{eq:base1}), indicating the data matrix $\mathbf{X}$ is appropriately standardized. The general construction of the prior can be written as,
 \begin{eqnarray}
X_l &\sim& N(0,\Sigma), \mbox{ }  \Sigma|D \sim \mathcal{IW}(b,D),  \mbox{ } l=1,\ldots,n, \nonumber \\
D &=& \mbox{diag}(d_1,\ldots,d_p), \mbox{ } d_k \sim \mathcal{G}_k(\bullet), \quad k=1,\ldots,p, \label{eq:modbase1}
\end{eqnarray}
where $\mathcal{IW}(b,D)$ denotes the Inverse-Wishart distribution with degrees of freedom $b$ and a diagonal scale matrix $D$. Here $\mathcal{G}_k(\bullet)$ denotes a general mixing distribution on the k-th diagonal element of the scale matrix of the Inverse-Wishart distribution, allowing adaptive shrinkage across different scales. By setting $\mathcal{G}_k(\bullet)$ to different mixing distributions, various types of shrinkage can be obtained - in this article, we shall explore shrinkage properties with respect to a specific choice of the mixing distribution, as elaborated in the sequel. 



Model (\ref{eq:modbase1}) relies on conjugate inverse Wishart prior on $\Sigma=\Omega^{-1}$, and varies from the traditional discrete mixture formulation (\ref{eq:base1}), in having a continuous prior on $\Sigma$. The traditional model (\ref{eq:base1}) constrains the support of $\Omega_G$ to the cone $M^{+}_G$ which depends on $G\in \mathcal{G}$, while the continuous prior in (\ref{eq:modbase1}) has an unconstrained support $M^+$ (the space of all positive definite matrices). Our choice of inverse Wishart formulation (\ref{eq:modbase1}) is based on both theoretic and computational considerations: (1) it induces a Gaussian distribution on the off-diagonals of $\Sigma^{-1}$ conditional on the precision diagonals (Lemma 1), a crucial factor in establishing graphical model selection consistency under the decision theoretic type approach proposed in Section 3, and (2) the associated conjugacy allows us to draw posterior samples of $\Sigma^{-1}$ as a whole, thus bypassing the computationally burdensome alternative of doing column-wise updates and enabling us to make significant computational gains - a key requirement in high dimensions.

Before proceeding further, it would be useful to define some notations. The covariance matrix is denoted as $ \Sigma \equiv \Sigma_p = \left( \begin{array}{cc}
\Sigma_{p-1,\bf{11}} &\sigma_{p,{\bf 2}1}         \\
      \sigma_{p,1{\bf 2}}  &\sigma_{p,pp} \end{array} \right), \mbox{ with }
      \quad  
      \Sigma_{k}^{-1} \equiv \Omega_k = \left( \begin{array}{cc}
      \Omega^{k-1}_{\bf{11}} &\omega_{k,{\bf 2}1}         \\
      \omega_{k,1{\bf 2}}  &\omega_{k,kk} \end{array}  \right)$, 
where $\Sigma_{p-1,\bf{11}}$  denotes the principal minor of dimension $p-1$ derived from the first $p-1$ rows and columns of $\Sigma$, and $\Omega^{k-1}_{\bf{11}}$ denotes the principal minor of dimension $k-1$ for $\Omega_k$. Let $\omega_{k,ij}$ denote the j-th element in the i-th row of $\Omega_{k}$, and note that $\omega_{k,kk}$ can be viewed as the residual precision for the regression $x_k|x_1,\ldots,x_{k-1},$ where $x_k$ denotes the generic k-th measurement, $i,j = 1,\ldots,k, \mbox{ } k=1,\ldots,p$.

Having defined such notations, we now state the following well known result as a Lemma, which serves as a first step towards understanding the regularization properties of the prior in (\ref{eq:modbase1}). The Lemma captures the distribution for elements in the last row of $\Sigma^{-1}\equiv\Omega_p$ conditional on $D$, and it is straightforward to adapt the result for any row.

{\noindent{\bf Lemma 1:}} For $\Sigma\sim \mathcal{IW}(b,\mbox{diag}(d_1,\ldots,d_p))$, we have
\begin{small}
$\pi(\omega_{p,pp}) = \mbox{Ga}\bigg(\frac{b+p-1}{2},\frac{d_p}{2}\bigg)$, and,  
$\pi(\omega_{p,1{\bf 2}}| \omega_{p,pp}) = \prod_{l=1}^{p-1} N(0,\omega_{p,pp}/d_l)$.
\end{small}

\bigskip

{\noindent \underline{\bf Choice of mixing distribution ($\mathcal{G}$):}} Lemma 1 shows us that the precision off-diagonals have a scale mixture representation under a prior on $d_1,\ldots,d_p$, and a careful choice of $\mathcal{G}_k(\bullet)$ might yield a prior on $\Sigma^{-1}$ with desirable shrinkage properties. In particular, the conditional distributions $\pi(\omega_{p,ij}|\omega_{p,ii}),i\ne j$ in Lemma 1 imply that a prior which assigns more mass to large values of $D$ is likely to induce greater shrinkage. However, imposing priors on $D$ by solely considering the conditionals $\pi(\omega_{p,ij}|\omega_{p,ii}),i\ne j$ may not adequate, as it does not immediately shed light on how the shrinkage will be induced through the joint prior on $\Sigma^{-1}$. We specify the following formulation, which leads to explicit expressions for the regularized prior after marginalizing out appropriate parameters
\begin{eqnarray}
d_k &\sim& \mbox{Inverse Gamma}(b/2 +1, \lambda_k^2/2) , \quad \lambda_k \sim \mbox{Ga}(a_{\lambda,k},b_{\lambda,k}), \quad k=1,\ldots,p,  \label{eq:scalemix}
 \end{eqnarray}
where $b$ is the degrees of freedom of the inverse Wishart in (\ref{eq:modbase1}). The hyperparameters $\bfl =(\lambda_1,\ldots,\lambda_p)$ control the shrinkage under the RIW approach - by proposing hyperpriors on $\bfl$ as in (\ref{eq:scalemix}), we can achieve a hierarchical specification which lets the data control the degree of shrinkage. We demonstrate the role of $\bfl$ in shrinkage in Figure 1, which plots the density of precision off-diagonals generated under the RIW formulation, with $\lambda_k=\lambda,k=1,\ldots,p,$ while fixing $b=p$. From the Figure, it is evident that higher values of $\lambda$ (and hence the diagonals of $D$) encourage greater shrinkage - we shall analytically explore the reason for this phenomenon in the next section. 

In the next section (2.2), we establish that under the inverse Gamma priors for $d_k,k=1,\ldots,p,$ as in (\ref{eq:scalemix}), the joint prior on $\Sigma$ after marginalizing out $D$ can be expressed in terms of a regularized prior on the elements of the Cholesky factor of $\Sigma^{-1}$. Hence we denote the prior in (\ref{eq:modbase1})-(\ref{eq:scalemix}) as Regularized Inverse Wishart (RIW) priors. Note that our specification is different from a conjugate Gamma specification on $d_k, k=1,\ldots,p$, which may not yield good estimates of the underlying graph due to lack of an initial regularization. Such an initial regularization provides important improvements over the unregularized counterpart, as demonstrated through simulations.

\vskip 12pt

{\noindent{\bf 2.2. Connections to Cholesky-based regularization}} \\
We now explicitly establish how our formulation in (\ref{eq:modbase1})-(\ref{eq:scalemix}) induces shrinkage in $\Sigma^{-1}=\Omega$ through equivalence with a Cholesky-based regularization approach. A modified Cholesky-based regularization procedures use the result (Newton, 1988, p. 359) that a symmetric matrix $\Sigma$ is positive definite if and only if there exists a unique diagonal matrix $V=(v_1,\ldots,v_p)$ with positive diagonal entries, and a unique lower triangular matrix $T$ having diagonal entries as 1, and off-diagonals satisfying  
\begin{small}
\begin{eqnarray*}
x_j = \sum_{k=1}^{j-1} (- t_{jk})x_k + \epsilon_j = \hat{x}_j + \epsilon_j, \quad j=2,\ldots,p,
\end{eqnarray*}
\end{small}
where $x_j$ is the generic representation for the j-th measurement, and $\epsilon_j $ has residual variance $v_j,j=2,\ldots,p,$ with $\epsilon_1 = x_1$. Then writing $\bfe = X - \hat{X}$, we have $\mbox{Var}(\bfe)=\mbox{Var}(TX)=T\Sigma T'$, which implies $\Sigma^{-1} = T'\mbox{diag}(v_1,\ldots,v_p)^{-1}T$. Using the above representation, the modified Cholesky-based regularization approaches induce shrinkage in $\Sigma^{-1}$ by imposing appropriate priors on the elements of $T,V$. 

In our case, the Gaussian structure in model (\ref{eq:modbase1}) allows the following equivalent representation as series of regressions
\begin{small}
\begin{eqnarray}
x_{k} = -\sum_{l=1}^{k-1} \omega^{-1}_{k,kk}\omega_{k,kl} x_l + \epsilon_{k}, \quad \epsilon_{k} \sim N(0,\omega^{-1}_{k,kk}),  k=1,\ldots,p,\label{eq:cholesky}
\end{eqnarray}
\end{small}
so that $\Sigma^{-1} = T'V^{-1}T$ with $V = \mbox{diag}(\omega^{-1}_{p,pp},\ldots,\omega^{-1}_{1,11})$ and $T$ being a lower triangular matrix having $t_{kl}= \omega^{-1}_{k,kk}\omega_{k,kl}, l<k$ and $t_{kk}=1, k=1,
\ldots,p$. Note that the ordering of equations in (\ref{eq:cholesky}) is not fixed and can vary under our approach. Clearly $\Sigma^{-1} = (T'V^{-1/2})(V^{-1/2}T)$, so that imposing a prior on the upper triangular Cholesky factor $T'V^{-1/2}$ will induce a corresponding prior on $\Sigma^{-1}$. Approaching the problem from the opposite direction, the prior on $\Sigma$ in (\ref{eq:modbase1})-(\ref{eq:scalemix}) will induce corresponding priors on the elements of the Cholesky factor $T'V^{-1/2}$, and hence on the regression coefficients in (\ref{eq:cholesky}). Theorem 1 shows that the induced prior after marginalizing out $D$ under formulation (\ref{eq:modbase1})-(\ref{eq:scalemix}) translates to a regularization prior involving a group Lasso type penalty on the regression coefficients in (\ref{eq:cholesky}). 

{\noindent{\bf Theorem 1:}} Under the prior defined in (\ref{eq:modbase1})-(\ref{eq:scalemix}), we can marginalize out $D$ to obtain $\pi(\Sigma|\bfl) \propto \prod_{k=1}^p\lambda^b_k(\omega_{k,kk})^{\frac{b+p-1}{2}}\exp\bigg(- \lambda_p\sqrt{|\omega_{p,pp}|}+\sum_{k=1}^{p-1}\lambda_k |\sqrt{ \sum_{l=0}^{p-k-1}\omega^{-1}_{p-l,p-l,p-l}\omega^2_{p-l,p-l,k}+\omega_{k,kk} }| \mbox{ } \bigg)$. \\
{\noindent{\bf Remark 1:}} The support of the prior defined by Theorem 1 is positive definite, which is evident from the fact that \begin{small} $P(\Sigma\in M^+|\bfl) = \int P(\Sigma\in M^+|D)d\pi(D|\bfl) = 1$, since $P(\Sigma\in M^+|D)=1$\end{small} almost surely with respect to $D$ by construction. \\
{\noindent{\bf Remark 2:}} It is straightforward to see that the prior is proper, since
\begin{small} 
\begin{eqnarray*}
\int_{\Sigma\in M^+} \pi(\Sigma|\bfl) \mbox{d}\Sigma = \int_{\Sigma\in M^+} \int \pi(\Sigma|D) d\pi(D|\bfl)\mbox{d}\Omega \stackrel{\mbox{Fubini}}{=}  \int \int_{\Sigma\in M^+}  \pi(\Sigma|D) \mbox{ d}\Sigma \mbox{ }d\pi(D|\bfl) = 1.
\end{eqnarray*}
\end{small}

To further explore the regularization aspects of the prior defined in Theorem 1, note that 
\begin{small}
\begin{eqnarray}
\exp\bigg(- \lambda_k |\sqrt{ \sum_{l=0}^{p-k-1}\omega^{-1}_{p-l,p-l,p-l}\omega^2_{p-l,p-l,k}+\omega_{k,kk} }| \mbox{ } \bigg)=\exp\bigg(- \lambda_{k} |\sqrt{ {\bf \tilde{t}}_k\mathcal{K}_{k}{\bf \tilde{t}}_k +\omega_{k,kk}}| \mbox{ } \bigg),  \label{eq:grplasso}
\end{eqnarray}
\end{small}
where  ${\bf \tilde{t}}_k$ corresponds to the first $k-1$ elements in the k-th column of $T'$ in (\ref{eq:cholesky}) and $\mathcal{K}_k = \mbox{diag}(\omega_{p,pp},\ldots,\omega_{k-1,k-1,k-1}),k=1,\ldots,p-1$. The exponent term in (\ref{eq:grplasso}) is reminiscent of the group Lasso penalty in the regression setting. The usual group lasso involving $J$ groups proceeds by solving the penalized regression
\begin{eqnarray*}
|| Y - \sum_{j=1}^J Z_j\gamma_j ||^2 + \mu \sum_{j=1}^J (\gamma_j'\mathcal{K}_j \gamma_j)^{1/2}, 
\end{eqnarray*}
where $\mu \ge 0$ is the regularization parameter and $\mathcal{K}_j'$s are called kernel matrices, and each $\gamma_j$ corresponds to the vector of regression coefficients for the j-th distinct group$,j=1,\ldots,J$. In our case, formulation (\ref{eq:modbase1})-(\ref{eq:scalemix}) is equivalent to fitting a series of $p-1$ regressions given by (\ref{eq:cholesky}) under the regularized prior in Theorem 1, with the order of the equations in (\ref{eq:cholesky}) being allowed to be arbitrary. In other words, our approach can be viewed as the Bayesian equivalent of the following penalized regression with group Lasso-type penalties
\begin{eqnarray*}
\sum_{k=2}^p \bigg(\omega_{k,kk}||X_{ck} - \sum_{j=1}^{k-1} (T')_j X_{cj} ||_2^2 + \lambda_{k} |\sqrt{ {\bf \tilde{t}}_k\mathcal{K}_{k}{\bf \tilde{t}}_k +\omega_{k,kk}}| \mbox{ } \bigg),
\end{eqnarray*}
where $(T')_j$ corresponds to the j-th row of $T'$ and $||\cdot||_2$ denotes the $L_2$ norm. In our case, the vector of regression coefficients for the k-th group consists of regression coefficients for $x_k$ in the regressions $x_l|x_1,\ldots,x_{l-1}, l>k$ in (\ref{eq:cholesky}). The k-th group has associated shrinkage parameter $\lambda_{k}$ and kernel matrix $\mathcal{K}_k$. It is easy to see that the prior in Theorem 1 is maximized with respect to $\lambda_k$ when $\lambda_k = \frac{b}{|\sqrt{ \sum_{l=0}^{p-k-1}\omega^{-1}_{p-l,p-l,p-l}\omega^2_{p-l,p-l,k}+\omega_{k,kk} }|}$, and similar conclusions hold for any element in $\bfl$. Thus a large value of $\frac{\lambda_k}{b}$ implies shrinkage for the elements of the Cholesky factor $T'V^{-1/2}$ and supports our earlier observation about Figure 1 that larger values of $\frac{\lambda_k}{b},k=1,\ldots,p,$ lead to greater shrinkage in $\Sigma^{-1}$. 


{\noindent \underline{\bf Order Invariance:}} Although Theorem 2 shows similarities between the proposed approach and Cholesky-based regularization, it is important to note a fundamental difference. In particular, most Cholesky-based regularization approaches are order dependent and induce shrinkage on $\Sigma^{-1}$ by specifying appropriate priors on the elements of the Cholesky factor, while fitting the series of $p-1$ regressions in (\ref{eq:cholesky}) in order to obtain posterior draws of $\Sigma^{-1}$. Instead, we propose conjugate inverse Wishart priors as in (\ref{eq:modbase1}) which are order invariant and naturally lead to shrinkage in $\Sigma^{-1}$ through a group Lasso type regularization on the elements of the Cholesky factor. The order invariance for the prior in Theorem 1 can be seen from the fact that \begin{small}$\pi(\Sigma_{\mathcal{P}}|\bfl) = \int \pi(\Sigma_{\mathcal{P}}|D) d\pi(D|\bfl) = \int\pi(\Sigma|D) d\pi(D|\bfl) = \pi(\Sigma|\bfl)$, \end{small} where $\pi(\Sigma_{\mathcal{P}}|D)=\pi(\Sigma|D)$ by construction in (\ref{eq:modbase1}), and $\Sigma_{\mathcal{P}}$ is the matrix obtained by permuting the rows and columns of $\Sigma$ in the order specified by permutation $\mathcal{P}$. If the ordering of equations in (\ref{eq:cholesky}) is changed according to the permutation $\mathcal{P}$, the prior in Theorem 1 would simply be expressed in terms of the elements of the Cholesky factor of $\Sigma^{-1}_{\mathcal{P}}$.

{\noindent{\bf 2.3. Posterior Computation}}\\
The MCMC sampler for the RIW proceeds using a straightforward fully Gibbs approach, using conjugacy to sample precision matrices as a whole. As mentioned previously, hyperparameters in the prior on $\lambda$ are important for determining the sparsity of the estimated graph. We specify $\lambda_k \sim \mbox{Ga}(a_{\lambda,k},1)$ for the RIW approach, where $a_{\lambda,1},\ldots,a_{\lambda,p}$ is an evenly spaced decreasing sequence from $n$ to $\max(n/2,p)$. Thus the prior mean of $\lambda_{k}$ is greater than $\lambda_{k'}$ for $k<k'$, which is specific to the ordering of equations in (\ref{eq:cholesky}) and needs to be adjusted for a different ordering. The reason for such a hyperparameter choice is due to Step 3 of posterior computation (below) in which we draw $\lambda_k$ from a Gamma posterior with shape parameter $b_{\lambda,k}+|\sqrt{ \sum_{l=0}^{p-k-1}\omega^{-1}_{p-l,p-l,p-l}\omega^2_{p-l,p-l,k}+\omega_{k,kk} }|,k\ge 2$. To induce appropriate shrinkage in the elements of $\bfl$s under the above posterior, we need $a_k>a_{k'}$ for $k<k'$ - such a choice works sufficiently well for a variety of simulation settings. We specify $b=3$ in our computations as in Jones et al. (2005). The MCMC alternates between the following steps. 

{\noindent{\em Step 1:}} Update $\Omega$ from the posterior  $\pi(\Omega|-) = \mbox{Wishart}(b+n,(D+\sum_{i=1}^n X_{i}X^T_{i})^{-1})$.\\
{\em Step 2:} Update $D$ using $\pi(d_p|-) = \mbox{Inv Gaussian}\bigg(\frac{\lambda_{p}}{\omega_{p,pp}},\lambda_p^2\bigg)$ and \\ $\pi(d_{p}|-) = \mbox{Inverse Gaussian}\bigg(\frac{\lambda_{k}}{ \sum_{l=0}^{p-k-1}\omega^{-1}_{p-l,p-l,p-l}\omega^2_{p-l,p-l,k}+\omega_{k,kk} }, \lambda^2_{k}\bigg), k=1,\ldots,p-1$. \\
{\em Step 3:} Update $\bfl$ using $\pi(\lambda_p|-) = \mbox{Ga}\bigg(b+a_{\lambda,p}+1,b_{\lambda,p}+\sqrt{|\omega_{p,pp}|}\bigg)$ and \\$\pi(\lambda_{k}|-) = \mbox{Ga}\bigg(b+a_{\lambda,k}+1,b_{\lambda,k}+\sqrt{  \sum_{l=0}^{p-k-1}\omega^{-1}_{p-l,p-l,p-l}\omega^2_{p-l,p-l,k}+\omega_{k,kk}} \bigg)$, k=1,\ldots,$p-1$. \\

\begin{center}{\bf 3. MODEL SELECTION AND CONSISTENCY}  \end{center}

{\noindent{\bf 3.1. Model selection}}\\
We now develop a post-MCMC fitting strategy for graphical model estimation, which assigns exact zeros to off-diagonals corresponding to absent edges by using a decision theoretic approach incorporating joint penalized credible regions. The proposed approach does not make any assumptions about the underlying graph structure, allowing for both decomposable and non-decomposable graphs. First, define the neighborhood of a node $i\in V$ as $\mbox{ne}_i = \{j\in V\setminus \{i\}: (i,j)\in E\}$ as in Meinshausen and B\"uhlmann (2006). Our approach uses connections with regression settings to perform neighborhood selection for each node in the graph, which are then subsequently combined to obtain estimates for the entire edge set.  In particular, the neighborhoods are estimated by using equivalent $L_0$ minimization based approaches in regression settings - in this paper, we adapt the approach proposed by Bondell and Reich (2012), to our context. We shall first briefly summarize the approach in Bondell and Reich (2012), and subsequently develop our approach for graphical model determination and discuss large-sample consistency properties in section 3.2.  

In particular, Bondell et al. (2012) first fit the full regression model 
\begin{eqnarray}
{\bf y} = {\bf Z}\bfb + \bfe,\quad  \epsilon_i \sim N(0,\sigma^2), \quad \beta_j \sim N(0,\sigma^2/\tau), \quad \sigma^2 \sim \pi(\sigma^2), \quad i=1,\ldots,n, j=1,\ldots,p, \label{eq:Bondell}
\end{eqnarray}
and subsequently perform model selection as a post-MCMC step by estimating an ordered sequence of models corresponding to a sequence of credible regions having probability content $1-\alpha$, with $\alpha\in(0,1)$ indexing the sequence. The model corresponding to a credible region $\mathcal{C}_\alpha$ with probability $1-\alpha$ in the sequence is obtained by finding a sparse solution for $\bfb$ by minimizing the $L_0$ norm $||\bfb||_0$ (i.e. the number of non-zero elements). In particular, they propose to solve 
\begin{eqnarray}
\tilde{\bfb} = \mbox{arg min}_{\bfb} ||\bfb ||_0, \quad  \mbox{ subject to } \bfb\in\mathcal{C}_\alpha = \{\bfb:(\bfb - \hat{\bfb})^T\hat{\Sigma}^{-1}(\bfb - \hat{\bfb}) \le C_\alpha \} , \label{eq:pencred}
\end{eqnarray}
where $C_\alpha$ is chosen to specify the $100(1-\alpha)\%$ joint credible interval $\mathcal{C}_\alpha$ and $\hat{\bfb},\hat{\Sigma}$ are the posterior mean and covariance of $\bfb$. 

Instead of directly solving (\ref{eq:pencred}) which depends explicitly on $\alpha$, Bondell et al. (2012) solve the equivalent Langrangian optimization problem 
\begin{eqnarray}
\tilde{\bfb} = \mbox{arg min}_{\bfb} (\bfb - \hat{\bfb})^T\hat{\Sigma}^{-1}(\bfb - \hat{\bfb}) + \Delta \sum_{j=1}^p |\hat{\beta}_j|^{-2}|\beta_j|, \label{eq:Lagrange}
\end{eqnarray}
where the proposed sequence of solutions corresponding to a sequence of credible regions is given as a function of $\Delta$. This results in a single parameter indexing the path, with there being a one-to-one correspondence between $\Delta, \alpha, C_\alpha$. Equation (\ref{eq:Lagrange}) can be solved using existing algorithms such as LARS (Efron et al., 2004) after some algebraic manipulations. For a fixed $\alpha/\Delta$, the solution to (\ref{eq:Lagrange}) corresponds to a particular model which excludes predictors corresponding to exact zeros under the $L_0$ norm minimization within the credible region $\mathcal{C}_\alpha$. 


{\noindent\underline{\bf Neighborhood Selection:}} In our graphical model selection context, denote ${\bfb}_k=\{\beta_{kj}=-\omega^{-1}_{p,kk}\omega_{p,kj}:j\ne k\}, k=1,\ldots,p$, and note that model (\ref{eq:modbase1}) admits the following representation for \begin{small} $\pi({\bfb}_k,\omega_{p,kk}|D, {\bf X})$ \end{small}
\begin{eqnarray}
x_{ik} &=& \sum_{j\ne k, j=1}^p \beta_{kj} x_{ij} + \epsilon_{ik}, \mbox{ } \epsilon_{ik}\sim N(0,\omega^{-1}_{p,kk}), \nonumber \\ \beta_{kj} &\sim& N(0,\omega^{-1}_{p,kk}/d_j), \quad \omega_{p,kk} \sim \mbox{Ga}\bigg(\frac{b+p-1}{2},\frac{d_k}{2}\bigg), \quad j\ne k, j=1,\ldots,p, \label{eq:Wreg} 
\end{eqnarray}
where the conditional normality of the regression coefficients in (\ref{eq:Wreg}) is specific to the inverse Wishart formulation, but is not guaranteed for arbitrary priors on $\Sigma$. After convergence of the MCMC, the posterior samples of $(\omega_{p,kk},{\bfb}_k)$ can be thought to be arising from the stationary distribution $\pi(\omega_{p,kk},{\bfb}_k|{\bf X})$ implied by (\ref{eq:Wreg}), with an additional prior specification for $D$ as in (\ref{eq:scalemix}) under the RIW approach. It is important to note here that the reverse does not hold - fitting the series of marginal models in (\ref{eq:Wreg}) will not result in posterior samples under the RIW approach, and more importantly may not result in a valid joint distribution (Scott and Carvalho, 2008). The prior on $D$ in (\ref{eq:scalemix}) results in shrinkage of unimportant $\beta_{kj}'s$ towards zero - such an initial regularization under RIW priors provides important improvements in neighborhood selection over the corresponding unregularized approach, as demonstrated in the simulation section. 


The posterior samples of $\Sigma^{-1}$ can be now directly used to obtain posterior realizations of $\bfb_k, k=1,\ldots,p,$ which can be thought as arising from fitting the marginal models (\ref{eq:Wreg}), but under a valid joint distribution as specified the RIW approach. Further, note that (\ref{eq:Wreg}) is very similar to model (\ref{eq:Bondell}) with ${\bf y}=X_{ck},{\bf Z}={\bf X}_{-k},\sigma^2 = \omega^{-1}_{p,kk},\bfb = {\bfb}_k$ for the k-th regression, where ${\bf X}_{-k}$ denotes {\bf X} without the k-th column. Hence, we can adapt the penalized joint credible regions approach to obtain a sparse solutions of $\bfb_k$ corresponding to level $\alpha$ as 
\begin{eqnarray}
\tilde{\bfb}^\alpha_k = \mbox{arg min}_{\bfb_k} ||\bfb_k ||_0, \mbox{ subject to } \bfb_k\in\mathcal{C}_\alpha = \{\bfb_k:(\bfb_k - \hat{\bfb_k})^T\hat{\Sigma_k}^{-1}(\bfb_k - \hat{\bfb_k}) \le C_\alpha \} , \label{eq:nbrhood}
\end{eqnarray}
where $\hat{\bfb_k}$ and $\hat{\Sigma_k}$ are the posterior mean and covariance of $\bfb_k$ respectively. The solution $\tilde{\bfb}^\alpha_k$ corresponds to a distinct estimated neighborhood $\hat{\mbox{ne}}_{k,\alpha}$=$\{l\in V: \tilde{\beta}^\alpha_{kl}\ne 0, l\ne k \}$ for node $k \in V$, since $\tilde{\beta}^\alpha_{kj}=0$ implies that the (k,j)-th precision matrix element is zero under the equivalence $\beta_{kj} = - \omega^{-1}_{p,kk}\omega_{p,kj},$ provided $\omega^{-1}_{p,kk}\ne 0$, $ j\ne k, k=1,\ldots,p$. The complexity of the proposed neighborhood selection approach for a fixed $\alpha/\Delta$ and for one node is $O(p^3)$, using equation (\ref{eq:nbrhood}) and complexity results of the LARS procedure.

{\noindent\underline{\bf Edge Set Estimation:}} Note that the edge set is defined as $E = \{(k,l): k\in \mbox{ne}_l \wedge l\in \mbox{ne}_k \}$. As we only consider undirected graphs, $k\in\mbox{ne}_l$ implies $l\in\mbox{ne}_k$, so that we also have $E=\{(k,l): k\in \mbox{ne}_l \vee l\in \mbox{ne}_k\}$ for our purposes. A particular estimate of the edge set corresponding to fixed $\alpha$ can be obtained by combining the neighborhoods for each node as $\hat{E}_{\alpha,\wedge} = \{(k,l): k\in \hat{\mbox{ne}}_{l,\alpha} \wedge l\in \hat{\mbox{ne}}_{k,\alpha}\}$. Alternatively, one can also define a less conservative estimate $\hat{E}_{\alpha,\vee} = \{(k,l): k\in \hat{\mbox{ne}}_{l,\alpha} \vee l\in \hat{\mbox{ne}}_{k,\alpha}\}$. Since the neighborhoods are estimated using posterior samples of $\Omega$ where $\omega_{p,kl}=\omega_{p,lk}, k\ne l$ and $\omega_{p,kk},k=1,\ldots,p,$ have more or less similar magnitudes after normalization of data, it is almost always the case that $\hat{E}_{\alpha,\wedge} = \hat{E}_{\alpha,\vee}$ in practice. Moreover it is shown in the next section that asymptotically both the estimates converge to the true edge set and are hence equal. We supress the second subscript and denote the estimated edge set with the generic notation $\hat{E}_{\alpha}, \alpha \in (0,1)$. 

{\noindent\underline{\bf Precision Matrix Estimation:}} Corresponding to a sequence of graphs, we also estimate a sequence of precision matrices, with the precision matrix corresponding to level $\alpha$ being computed as $\hat{\Omega}_{\hat{E}_{\alpha}}=\hat{\Omega}\otimes ADJ_{\alpha}$, where $\hat{\Omega}$ is the posterior mean of the MCMC samples, $ADJ_{\alpha}$ is the adjacency matrix corresponding to the edge set $\hat{E}_{\alpha}$ and $\otimes$ denotes Hadamard product. Thus $\hat{\Omega}_{\hat{E}_{\alpha}}$ has exact zeros corresponding to absent edges in $\hat{E}_{\alpha}$ and non-zero entries estimated as the posterior mean from the MCMC samples. Such an estimate is asymptotically consistent as shown in the next section, and hence positive definite. In practice, since the model selection approach assigns to near zero elements of the precision matrix, $\hat{\Omega}_{\hat{E}_{\alpha}}$ is essentially almost always positive definite for the numerical examples we considered.

{\noindent{\bf 3.2. Selection Consistency}}

In this section, we establish that the proposed model selection approach involving penalized joint credible regions leads to consistent neighborhood selection under some suitable assumptions. Suppose that for a given sample of size $n$, we estimate the neighborhood corresponding to level $\alpha_n$ in the ordered sequence, and denote the corresponding estimated neighborhood for the $k$-th node as $\hat{ne}_{k,n}$. By choosing $\alpha_n$ such that $1-\alpha_n \to 1$ as $n\to \infty$, (i.e. the coverage increases with $n$), we show that under such a choice the probability of the neighborhood of node $k$ equaling the true neighborhood $\mbox{ne}_{k0}$ goes to 1 as $n\to \infty$. 


For a sample size $n$, let the credible region for $\bfb_k$ with content $1-\alpha_n$ be $\mathcal{C}_{n,k} = \{\bfb_k:(\bfb_k - \hat{\bfb}_k)^T\hat{\Sigma}_k^{-1}(\bfb_k - \hat{\bfb}_k) \le C_{n} \}$. Using Lemma 1 in Bondell et al. (2012), $C_{n}\to \infty$ implies $1-\alpha_n \to 1$, implying a one-to-one correspondence between $\alpha_n$ and $C_{n}$. Suppose $p$ is fixed and consider the following assumptions: 

{\noindent(A1)} The true model is $X \sim N(0,\Omega_{E_0}^{-1})$, where $\Omega_{E_0}=(\omega_{0,ij})_{i,j=1}^p$ has exact zeros for off-diagonals corresponding to absent edges in $E_0$. Here, $E_0$ is the true edge set corresponding to an undirected graph and having true neighborhood $\mbox{ne}_{k0}$ for node $k,\mbox{ }k=1,\ldots,p$. \\
{\noindent(A2)} $\Omega_{E_0}$ is positive definite with $c_1/\sqrt{n}<|\omega_{0,ij}|< c_2$ for finite and positive constants $c_1,c_2,$ for all $\{\omega_{0,ij}:(i,j)\in E_0 \}$. 

The following Theorem establishes neighborhood selection consistency.

{\noindent{\bf Theorem 2:}} Under assumptions (A1), (A2), if we choose a sequence of credible regions $\mathcal{C}_{n,k}$ such that $C_n\to\infty$ and $n^{-1}C_n\to 0$, then $P(\hat{\mbox{ne}}_{k,n}=\mbox{ne}_{k0})\to 1$ as $n\to \infty$, $k=1,\ldots,p$. 

{\noindent{\bf Remark 3:}} Theorem 2 holds for any proper mixing distribution $d_k \sim \mathcal{G}_k(\bullet)$, as well as for non-stochastic $d_k=o(n),k=1,\ldots,p$. 

The proof of the above Theorem is provided in the Appendix. Let the estimated generic edge set for level $\alpha_n$ be denoted as $\hat{E}_{n}$ with the corresponding estimated precision matrix as $\hat{\Omega}_{\hat{E}_n}$. The following result holds for both $\hat{E}_{n,\wedge}$ and $\hat{E}_{n,\vee}$. 

{\noindent{\bf Corollary 1:}} Suppose Theorem 2 holds. Then $P(\hat{E}_{n}=E_{0})\to 1$ and $P(\hat{\Omega}_{\hat{E}_n}=\Omega_{E_0})\to 1$ as $n\to \infty$. 

The first part of the above Corollary follows since conditional independence structure of a multivariate normal can be consistently estimated by combining the neighborhood estimates of all variables (Meinshausen and B\"uhlmann, 2006). The edge set estimates $\hat{E}_{n,\wedge}$ and $\hat{E}_{n, \vee}$ converge asymptotically when the truth is a Gaussian graphical model as in (A1), due to consistent selection of the neighborhoods in Theorem 2. The proof for the second part of Corollary 1 is in the Appendix. Since $\Omega_{E_0}$ is positive definite by assumption, $\hat{\Omega}_{\hat{E}_n}$ is asymptotically positive definite.

{\noindent{\bf 3.3. Edge selection based on false discovery rates}}

While, the above approach yields an ordering of graphs which lead to asymptotic consistency, it is often of interest to report a single point estimate of a graph that is best supported by the data. Usual Bayesian methods obtain such a graphical estimate by including edges have a posterior probability $>0.5$, i.e. median probability model (Barbieri and Berger, 2004) or reporting the graph having the highest log-likelihood while frequentist approaches minimize some BIC-type criteria, the latter often leading to sparse estimated graphs (as demonstrated in simulations). Instead, we propose here an approach based on controlling false discovery rates (FDR) which includes a natural multiplicity correction and can directly control the level of sparsity in edge selection.

First note that the edges that are strongly supported by the data will likely appear in most of the ordered sequence of graphs, whereas other edges with weaker evidence may appear less often. We first compute a pseudo posterior inclusion probability matrix $P= (P_{ij})$ by computing the proportion of times each edge is included in the ordering of graphs based on the sequence of penalty parameters $\{\Delta_m\}_{m=1}^R$, where $R$ is the chosen number of penalty parameters. Then $1-P_{ij}$ can be considered akin to Bayesian q-values, or estimates of the ``local false discovery rate" (Storey et al., 2003; Newton et al., 2004), as they measure the probability of a false positive if the (i,j)-th edge is called a `discovery' or is significant. Given a desired global FDR bound $\eta\in (0,1)$ (implying that we expect only $100\eta\%$ of the edges that are
declared significant are in fact false positives), one can determine a threshold $c_\eta$ which flags the set of important edges as $\hat{E}_\eta = \{(i,j): P_{ij}\ge c_\eta \}$. This yields a point estimate of the graph. 

We adapt the approaches in Morris et al. (2008) and Baladandayuthapani et al., (2010) to our context, in order to determine the significance threshold $c_\eta$ by controlling the average Bayesian FDR. Let vec($P$) be the vectorized upper triangular matrix of $P$ excluding diagonals, containing the pseudo posterior inclusion probabilities of the edges stacked columnwise. We first sort vec($P$) in descending order to yield the sorted vector vec($\tilde{P}$)=$\{\tilde{P}_{k}, k=1,\ldots, p(p-1)/2 \}$. Then we can estimate $c_\eta$ as the $\zeta$-th entry of vec($\tilde{P}$), where $\zeta = \max\{j^*:\frac{1}{j^*}\sum_{k=1}^{j^*} \tilde{P}_{k} \le \eta \}$, and a lower value of $c_\eta$ leads to sparser graphs, while simultaneously controlling the false discovery rate at a pre-specified level. 

\vskip 12pt

\begin{center} {\bf 4. SIMULATION STUDIES }   \end{center}

We present simulation studies for two data generating models and different ($n,p$) combinations, while comparing our approach (RIW) to discrete mixture based approaches such as the hyper inverse Wishart approach using reversible jump MCMC as in Bhadra et al. (2012) but excluding predictors (HIW), continuous shrinkage approaches such as the Bayesian graphical lasso (BGLA) and Bayesian adaptive graphical lasso (BGAD) with default hyperparameter values as implemented in Wang (2012) (Matlab code available in their supplementary materials), and the unregularized inverse Wishart prior (IW) which has the same formulation as in (\ref{eq:modbase1}), but with $D=dI_p$ and $d\sim Ga(1,1)$, and subsequently estimating the graph structure using the same approach as in section 3. We also compare our approach to the frequentist graphical lasso (GLASSO) (Friedman et al., 2008) as implemented in the Matlab Glasso code available at www.stanford.edu/~tibs/glasso. For RIW, BGLA and BGAD procedures, 15000 MCMC iterations with a burn in of 5000 was used, while 100000 iterations with burn in of 10000 was used for HIW. The initial adjacency matrix for the HIW corresponds to a null graph, as in Bhadra et al. (2012). 

We consider two cases for data generation, with each case having 50 replicates. \\
{\bf Case I:} A fractional Gaussian noise process having covariance elements 
\begin{eqnarray*}
\sigma_{ij} = \frac{1}{2} \left[||i - j| + 1|^{2H} - 2|i-j|^{2H} + ||i-j| - 1|^{2H}\right],
\end{eqnarray*}
where $H\in [0.5,1]$ is the Hurst parameter, and choosen to be $H=0.7$ as in Banerjee et al. (2012). Data was generated for $(n,p)=(300,100),(400,200), (500,100), (500,200)$, and $(700,500)$. For $(n,p)=(700,500)$, we could only compare the performance of RIW and GLASSO, as it was not possible to obtain results for any of the other Bayesian competitors due to unrealistic computational burden. 

{\noindent \bf Case II:} For this case, we generate data emulating a real data application, where mRNA expression levels for 49 ($p$) genes are available for 241 ($n$) subjects. From prior biological evidence, it is known that these 49 genes have underlying connections between them, so that they can be said to lie on a graph - these aspects will be described in more detail in section 5.1. We first fit the RIW model to this data, and obtain an estimated graph having edge set $\hat{E}$ for $\Delta=0.05$ and the corresponding positive definite precision matrix ($\hat{\Omega}_{\hat{E}}$) with exact zeros corresponding to absent edges in $\hat{E}$. The resulting graph has 631 edges and is shown in Figure 4 (a). Subsequently, we generate data under a zero mean Gaussian graphical model with covariance $\hat{\Omega}^{-1}_{\hat{E}}$, for $n=200,300$. 

As the credible set level $\alpha$ is varied, an ordered set of graphs is created under the RIW approach. Similarly, as the penalty parameter in Glasso and the posterior inclusion threshold in HIW is varied, ordered sets are created. Following Wang (2012), we can create ordered graphs for BGLA and BGAD by including the edge $(i,j)$ if  $\frac{\tilde{\rho}_{ij}}{E_g(\rho_{ij}|{\bf X})}> q_m, q_m\in(0,1)$, and excluding others. Here the numerator and denominator are the posterior mean of partial correlations under his approach and a reference distribution $\mbox{Wishart}(3,I_p)$. To assess the performances of different approaches, we first consider the induced ordering of graphs. 

For each point in the ordering, we denote the true positives (TP), true negatives (TN), false positives (FP), and false negatives (FN), as those edges that are correctly included, correctly excluded, incorrectly included, and incorrectly excluded respectively. For each point on the ordering, we obtain specificity and sensitivity as $SP = \frac{TN}{TN+FP}, \quad SE = \frac{TP}{TP + FN}$, where the denominators in SP and SE correspond to the total number of absent and present edges respectively. Thus specificity = 1- False Positive Rate (FPR), and sensitivity can be considered as the power for detecting correct edges. 


\underline{\bf{True graph recovery:}} We look at the area under the Receiver-Operating Characteristic (ROC) curve to compare performance of graph rankings. ROC curves plot the sensitivity versus FPR or 1-specificity, and is related to the trade-off between type-I error and power. Instead of one ROC curve, we look at a series of ROC curves, with the m-th element in the series corresponding to the true edge set defined by including edges having absolute partial correlations above the m-th threshold $c_m$. By varying $c_m \in [0.005,0.26]$, we obtain a series of true edge sets with a higher value of $c_m$ implying a lower number of edges corresponding to weak partial correlations. The series of areas under the ROC curve for different true edge sets and $(n,p)$ combinations are plotted in Figure 2. 

A general examination of Figure 2 reveals that for all approaches, the area under the ROC curve is lower under lower values of $c_m$, but increases as the threshold is increased. This is to be expected, as it is more difficult to detect edges corresponding to near zero partial correlations versus edges corresponding to stronger partial correlations. We see that for Case I, RIW has highest area under the curve corresponding to low values of $c_m$, but as $c_m$ is increased, RIW, GLASSO and IW have similar performance with area under the curve close to one. In contrast other competing approaches have substantially poorer performance, with the area under the curve not nearing one even for large values of $c_m$.

On the other hand for Case II, it is clear that RIW dominates all other approaches, while HIW has a substantially improved performance in contrast to the high dimensional Case I. Continuous shrinkage based approaches such as BGLA and BGAD perform comparatively poorly, with BGLA yielding substantially poorer results for $n=300$. We report the area under the ROC curve for true edge set ES1 corresponding to moderately strong absolute partial correlations ($c_m = 0.1$), and for true edge set ES005 corresponding to low absolute partial correlation ($c_m = 0.005$) in Tables 1.1 and 1.2. 

To examine the reason for low area under the curve for competing approaches, we plot the ROC curves in Figure 3 when the true edge set is ES1. For Case I, we see that the ROC curves for BGLA, BGAD and HIW, are uniformly dominated by RIW and GLASSO, with the ROC curve under HIW being a segment-wise straight line. This results from either extremely high or low sensitivity values (with no intermediate values) under different thresholds for the edge specific posterior inclusion probabilities and seems to be associated with the difficulties of reversible jump MCMC in efficiently exploring the graph space. For Case II, it is evident that the ROC curve under RIW uniformly dominates all other curves except HIW, with the latter demonstrating a much improved performance compared to the high dimensional Case I. It is clear that under Case II when $n=300$, the ROC curves for BGLA and BGAD rapidly taper towards the origin, thus exhibiting remarkably low sensitivity for high specificity levels compared to other approaches, which leads to low area under the curve.

The preceeding results throw light on the performance under the induced ordering of graphs. We now examine point estimates for the graph under different approaches in Tables 2.1 and 2.2. We use $\eta=0.2$ for the FDR based thresholding approach for RIW and IW approaches. For HIW approach, we follow the usual procedure of estimating the graph by only including edges having marginal inclusion probabilities $>0.5$. For BGLA and BGAD, we can obtain a point estimate for the edge set by only including a particular edge $(i,j)$ if $\frac{\tilde{\rho}_{ij}}{E_g(\rho_{ij}|{\bf X})}>0.5$, as in Wang (2012). For the Glasso approach, the optimal graph is obtained by minimizing the BIC criteria used in Yuan and Lin (2007)
\begin{small}
\begin{eqnarray*}
\mbox{BIC}(\Delta) = n \times \bigg(-\log(|\hat{\Sigma}^{-1}_{\Delta}|) + \mbox{trace}(\hat{\Sigma}^{-1}_{\Delta} ({\bf X}^T{\bf X})/n) \bigg) + \frac{\log(n)}{n}\times \#\{(i,j): 1\le i\le j \le p, \hat{\omega}^{\Delta}_{ij}\ne 0 \},  
\end{eqnarray*}
\end{small}
where $\Sigma^{-1}_\Delta=(\hat{\omega}^{\Delta}_{ij})$ denotes the estimated precision matrix for penalty parameter $\Delta$.

Tables 2.1-2.2 report sensitivity and specificity under estimated graph best supported by the data, for true edge sets ES1 and ES005. Under Case I, we see that none of the approaches is a clear winner - however, the RIW approach is seen to have higher specificity for all cases and higher sensitivity for true edge set ES1 compared to BGLA. RIW also demonstrates improvements in both specificity and sensitivity over BGAD in several cases in Table 2.1. For Case II, the sensitivity under RIW is always within the highest two values reported under any approach, while having reasonably high specificity levels as well. GLASSO seems to report sparse graphs with low sensitivity under the BIC criteria, even for true edge set ES1. On the other hand, the results under the HIW approach seem to vary widely under different choices of the adjacency matrix. For Case I when $(n,p)=(300,100)$, the specificity and sensitivity for HIW was 45, 57, for true edge set ES005 and 45,89, for ES1 under an initial adjacency corresponding to a complete graph - these results are very different compared to those reported in Table 2.1 under an initial null adjacency matrix.

The difference in results under different initial adjacency matrices is indicative of a larger issue - the instability of HIW results for higher dimensions under finite runs of the MCMC. To demonstrate this, Figure 5 shows a histogram of the standard errors of the posterior inclusion probabilities for the edges under the HIW approach for Case I when $(n,p)=(300,100)$, over different replicates. It is evident that some of these standard errors can be as high 0.4, thereby potentially resulting in unstable estimates of edge sets across replicates. Similar unstable behavior was reported by Scott and Carvalho (2008) for Metropolis based approaches under discrete mixture priors. Combined with the segment-wise straight line ROC curves in Case I associated with difficulties of the reversible jump MCMC in efficiently exploring the graph space, it is evident that applicability of HIW (and in general most discrete mixture approaches) can become increasingly challenging for higher dimensions under finite runs of MCMC. 

For the high dimensional case $(n,p)=(700,500)$ under Case I, Table 3 reports the area under the ROC curve and sensitivity and specificity values under the optimal graph for RIW and GLASSO. It can be seen that RIW has higher area under the curve under the true edge set ES005, while both approaches have area = 1 under the true edge set ES1. In terms of performance under the optimal graph, the GLASSO seems to do marginally better with slightly higher sensitivity levels. Given the fact that it was not possible to apply BGLA, BGAD and HIW, due to an unrealistic computational burden, this example highlights the advantage of our approach over Bayesian alternatives in high dimensions. 

\underline{\bf{Detection of hub nodes:}} For Case II, we also look at the performance of different approaches for detecting the neighbors of `hub' nodes, which are defined as important nodes having a high number of connections ($>30$ neighbors for our example), and are important from a biological perspective as discussed in Section 5. The number of neighbors for each hub node is called it's degree. We examine how well different approaches compare in terms of estimating the degree for each hub node, for true edge sets ES1 and ES005. Figures 4(c)-(d) plot the true and estimated degrees under the optimal graph estimated by different approaches, with the x-axis corresponding to different hub nodes. We see that RIW and BGLA perform better than the other approaches in estimating the degree, with the GLASSO having relatively poor performance due the estimated graph being parsimonious under the BIC criteria. The difference between true and estimated degrees can be high due to presence of weakly related edges for ES005, but the difference comes down for ES1 which has edges corresponding to moderate or strong absolute partial correlations.


\underline{\bf{Computational Efficiency:}} An important aspect of any Bayesian procedure is computational efficiency. Table 4 reports the computation time in cpu seconds for the Bayesian approaches in Case I for different $(n,p)$ combinations. It is clear that the RIW approach is several times faster than the continuous shrinkage based approaches BGLA and BGAD, and is scalable to higher dimensions. We conjecture that the reason for these competing approaches to be computationally intensive is due to the requirement of sampling $\Omega$ by doing column-wise updates which becomes increasingly burdensome as $p$ increases. On the other hand, the computation time under the HIW approach per iteration is slower but comparable to that under the RIW - however the HIW requires increasingly longer (and perhaps infeasible) MCMC runs to attain meaningful results as $p$ increases, with the net computation time exploding.  

\underline{\bf{Conclusions:}} From the simulation results, it is clear that the RIW approach (1) outperforms competing Bayesian approaches in terms of true graph recovery in high dimensions, (2) has demonstrably better computational efficiency over other competing Bayesian approaches, with the latter approaches quickly becoming computationally infeasible as $p$ increases, and (3) in terms of area under the ROC curve, exhibits improvements over GLASSO under Case I for true edge sets defined by lower thresholds of absolute partial correlations, while dominating GLASSO when data is generated emulating a real data application in Case II.

\vskip 12pt

\begin{center}{\bf 5. APPLICATION TO CANCER GENOMICS} \end{center}


We illustrate our methods using a glioblastoma multiforme (GBM) genomics  dataset collected by The Cancer Genome Atlas (TCGA) network. GBM was one of the first cancers evaluated by the TCGA and has various molecular measurements on over 500 samples, that include gene expression (mRNA) and microRNA(miRNA) expression among many others (see http://tcga-data.nci.nih.gov/tcga). mRNAs and miRNAs play complimentary roles in disease progression and development and were recently found to be associated to many cancers especially GBM (Tang et al., 2013). The key scientific questions we address are bi-fold: find important mRNA and miRNA regulatory networks/connections and to detect ``hub" components in these networks that might point to major drivers in the etiology of GBM development.  

We analyze two subsetted data sets - one involving 49 mRNA expressions (moderate $p$) for 241 GBM patients, and another having 250 miRNA expressions (large $p$) for 280 GBM patients. The mRNA/genes were selected from core pathways implicated in GBM such as receptor tyrosine kinase (RTK), phosphatidylinositol-3-OH kinase (PI3K) and etinoblastoma (RB) pathways (McLendon, R. et al., 2008), and the data was directly obtained from the TCGA website. The miRNA data set initially consisted of 538 miRNA expressions and the survival time for 280 subjects. As a pre-processing step, we fit a univariate Cox proportional hazards model for each of the miRNAs, and subsequently selected 250 miRNAs having significant effects on survival times, based on ranking of $p$-values to focus on the top miRNAs. As the underlying graph is expected to be sparse in both the applications, we use $\eta=0.1$ in our FDR based approach for edge selection and we summarize our major findings below.

{\noindent{\bf \underline{Gene(mRNA) regulatory networks:}}
The hub nodes along with the number of neighbors are listed in Table 5.1, and the estimated graph is shown in Figure 6. The estimated graph had 6 hub nodes each having greater than 8 neighbors, while one node did not have any neighbors.  Some of the highly connected genes such as PI3KC2G (14 connections), EGFR (9 connections) and CDKN2A (9 connections) have been previously shown to be associated with glioblastoma (Dong et al., 2010; Wong et al., 1992; Herman et al., 1995).  We further explored the biological implications of our results using a pathway analysis of genes  using  ingenuity pathway analysis (IPA version 16542223) which examines partial correlations under the RIW approach to identify important functional pathways implicated in literature. The IPA analysis identified a number of enriched pathways including; glioma, GBM, PTEN signaling and other molecular mechanisms in known in cancer. This is so since most of these genes encode protein critical to cellular functions such as cancer, DNA recombination, and repair, cellular development, cell cycle and connective tissue development which may be attributed to their highly connected nature. \\

 {\noindent \underline{\bf MicroRNA regulatory networks:}} The estimated miRNA graph had 9 hub nodes each having greater than 10 neighbors, while 107 nodes did not have any neighbors. The hub nodes along with the number of neighbors are listed in Table 5.2, and the estimated graph is shown in Figure 6. Analogous to gene expression a similar analysis of the miRNA with at least 4 neighbors (based on partial correlations under RIW approach) using IPA suggest they are critical for various cellular processes in cancer progression. The selected molecules modulate important transcription factors and signaling molecules including genes such as MYC, CCLE1 and CLDND1 which have been shown to be associated with cancer, inflammatory response and connective tissue disorders. This concurs with studies that have indicated significant modulation of listed miRs such as miR-106 (26 connections), miR-184 (15 connections) and miR-let-7a (26 connections), with glioblastoma (Wang et al., 2012; Malzkorn et al., 2010; Lee et al., 2011).

\vskip 12pt

\begin{center} {\bf 6. DISCUSSION }  \end{center} 

We have proposed a novel Bayesian graphical model selection approach that overcomes several difficulties of existing Bayesian approaches. The proposed approach is shown to be selection consistenct in recovering the true graphical model, and is scalable to higher dimensions, thus providing a theoretically justified Bayes graphical model selection approach which can address high dimensional settings. 

\vskip 12pt

\begin{center}{\bf APPENDIX } \end{center} 

{\noindent {\bf Proof of Theorem 1:}} Without loss of generality, let $\mathcal{P}=\{p,,p-1, \ldots,2,1\}$ denote the ordering of equations in (\ref{eq:cholesky}), and let $\Omega_{\mathcal{P}}=Q_{\mathcal{P}}\Omega Q'_{\mathcal{P}}$, with $Q_{\mathcal{P}}$ representing the permutation matrix corresponding to $\mathcal{P}$. Then, using the arguments in section 2.2, we have  $ \Omega_{\mathcal{P}}= T'V^{-1}T$ where \begin{small}$V^{-1/2}T = \left( \begin{array}{cccc}
\omega^{1/2}_{p,pp}  &\omega^{-1/2}_{p,pp}\omega_{p,p,p-1} &\omega^{-1/2}_{p,pp}\omega_{p,p,p-2} \ldots      &\omega^{-1/2}_{p,pp}\omega_{p,p,1} \\
0 &\omega^{1/2}_{p-1,p-1,p-1} &\omega^{-1/2}_{p-1,p-1,p-1}\omega_{p-1,p-1,p-2} \ldots &\omega^{-1/2}_{p-1,p-1,p-1}\omega_{p-1,p-1,1}\\
\vdots &\vdots &\vdots &\vdots\\
0      &0      &0      &\omega^{1/2}_{1,11}\end{array} \right)$ \end{small}.\\

Equating the trace of $D\Omega_{\mathcal{P}}$ with that of $DT'V^{-1}T$, we have,
\begin{eqnarray*}
\mbox{trace}(D\Omega_{\mathcal{P}}) = d_p\omega_{p,pp} + \sum_{k=1}^{p-1}d_k \bigg( \sum_{l=0}^{p-k-1}\omega^{-1}_{p-l,p-l,p-l}\omega^2_{p-l,p-l,k}+\omega_{k,kk} \bigg)
\end{eqnarray*}
Also note that $\mbox{det}(\Omega_{\mathcal{P}})= \prod_{k=1}^p\omega_{k,kk}$. Using the form of the inverse Wishart density, these facts imply 
\begin{eqnarray*}
f(\Omega_{\mathcal{P}}|D) \propto \prod_{k=1}^p d^{b/2}_k \omega^{\frac{b+p-1}{2}}_{k,kk}\exp\bigg(-\frac{1}{2}d_p\omega_{p,pp} -\frac{1}{2} \sum_{k=1}^{p-1}d_k \left[ \sum_{l=0}^{p-k-1}\omega^{-1}_{p-l,p-l,p-l}\omega^2_{p-l,p-l,k}+\omega_{k,kk} \right]\mbox{ } \bigg)
\end{eqnarray*}
Then writing $\tau_k=1/d_k, k=1,\ldots,p,$ we have $f(\Omega_{\mathcal{P}}|\bfl) = \int f(\Omega|D)d\pi(D)$
\begin{small}
\begin{eqnarray*}
\propto \int \prod_{k=1}^p\frac{1}{\sqrt{\tau_k}}\omega^{\frac{b+p-1}{2}}_{k,kk}\exp\bigg(-\frac{1}{2}\left[\frac{\omega_{p,pp}}{\tau_p} + \sum_{k=1}^{p-1}\frac{ \sum_{l=0}^{p-k-1}\omega^{-1}_{p-l,p-l,p-l}\omega^2_{p-l,p-l,k}+\omega_{k,kk} }{\tau_k}\right] -\frac{1}{2}\sum_{k=1}^p\lambda^2_k\tau_k\bigg)d\tau_1,\ldots, d\tau_p.
\end{eqnarray*}
\end{small}
Using the scale mixture representation of the Laplace distribution, we have the result.

\vskip 12pt

{\noindent{\bf Proof of Theorem 2:}} The proof uses sufficiency conditions of Theorem 1 in Bondell and Reich (2012), a result which we describe now. In context of fitting regression model (\ref{eq:Bondell}), suppose $\tilde{\bfb}^{\alpha_n}$ is the solution to (\ref{eq:pencred}) with respect to the credible region having probability content $1-\alpha_n$. Denote the estimated set of non-zero coefficients in (\ref{eq:Bondell}) as $A_n=\{j:\tilde{\beta}_j^{\alpha_n}\ne 0 \}$ and let the true set be $A=\{j: \beta_j^0 \ne 0 \}$, with $\bfb_0$ being the vector of true regression coefficients. Consider a sequence of credible sets $\mathcal{C}_n=\{\bfb:(\bfb - \hat{\bfb})^T\hat{\Sigma}^{-1}(\bfb - \hat{\bfb}) \le C_{\alpha_n} \}$ such that $P(\mathcal{C}_n)= 1-\alpha_n$, where the coverage $1-\alpha_n$ increases with $n$. Assume the following regularity conditions:\\
(B1) The true error terms are i.i.d. with mean zero and finite variance. \\
(B2) The matrix $Z^TZ/n\to Q,$ where $Q$ is positive definite. \\
(B3) The prior precision $\tau$ satisfies $\tau=o(n)$. \\
(B4) $\min\{|\beta_j^0|: j\in A \} > c_1/\sqrt{n},$ for some $c_1>0$. \\
{\noindent \bf {Theorem 3:}} (Bondell et al., (2012)) Under conditions (B1) - (B4), if $C_n\to\infty$ and $n^{-1}C_n\to 0$, then $P(A_n=A_0)\to 1$. 

For our case, consider the conditional regression $x_k|x_1,\ldots,x_{k-1},x_{k+1},\ldots,x_p$ as in equation (\ref{eq:Wreg}). As argued previously in section 3.1, the posterior samples of $\pi(\bfb_k,\omega_{k,kk}|{\bf X})$ under the RIW approach can be equivalently thought as arising from fitting the model (\ref{eq:Wreg}), $k=1,\ldots,p$. Under assumption (A1) of Theorem 2, the truth is a Gaussian graphical model, and hence admits the following set of true conditional regressions
\begin{eqnarray}
x_{ik} = \sum_{j\ne k,j=1}^p \beta^0_{kj} x_{ij} + \epsilon^0_{ik}, \quad \epsilon^0_{ik} \sim N(0,\omega^{-1}_{0,kk}),\mbox{ } i=1,\ldots,n, \label{eq:truereg}
\end{eqnarray}
where $\beta^0_{kj}$ are the true regression coefficients with $ \beta^0_{kj}= 0$ when $\omega_{0,kj}=0\mbox{ } (j\ne k)$ corresponding to absent edges in $E_0$. We will show that the sufficiency conditions (B1)-(B4) in Theorem 3 hold for our case for each regression in (\ref{eq:Wreg}), which will help prove Theorem 2. 

Without loss of generality, we will first establish consistency for the k-th conditional regression $X_{ck}|{\bf X}_{-k}$ in (\ref{eq:Wreg}), and subsequently marginalize out ${\bf X}_{-k}$ to obtain our result. Under the representation (\ref{eq:truereg}) for the k-th true regression and assumption (A2), condition (B1) is clearly satisfied. Further for fixed $p$, ${\bf X}^T{\bf X}/n \stackrel{a.s.}{\rightarrow} \Omega^{-1}_{E_0}$, so that ${\bf X}^T{\bf X}/n $ is positive definite as $n\to \infty$ (under assumption (A2)) except on a set of measure zero. Substituting ${\bf X}_{-k}$ for ${\bf Z}$, the above fact ensures that condition (B2) in Theorem 3 is satisfied for the conditional regression $X_{ck}|{\bf X}_{-k}$, for all ${\bf X}_{-k}$ with positive probability. Further for all $(k,j)\in E_0$, $\beta^0_{kj} = \omega^0_{k,kj}/\omega^0_{k,kk}$, so that condition (B4) is satisfied under assumption (A2).

Instead of having a fixed prior precision $\tau$ as in Theorem 3, the prior precision for $\beta_{kj}$ is $d_j,j\ne k$, with $\pi(D)$ defined in (\ref{eq:scalemix}). The posterior mean and variance of $\bfb_k$ is given by 
\begin{small}
\begin{eqnarray*}
\hat{\bfb}_k\approx \frac{1}{M-B}\sum_{m=B+1}^M({\bf X}^T_{-k}{\bf X}_{-k} + D^m_{-k})^{-1}{\bf X}^T_{-k}X_{ck},\quad  \hat{\Sigma}_{\bfb_{k}} \approx \frac{1}{M-B}\sum_{m=B+1}^M \tilde{s}_k({\bf X}^T_{-k}{\bf X}_{-k} + D^m_{-k})^{-1}, 
\end{eqnarray*}
\end{small}
where $M$ is total number of MCMC iterations, $B$ is burn-in, and $D^m_{-k}$ is the m-th MCMC sample for a diagonal matrix with diagonals $(d_1,\ldots,d_{k-1},d_{k+1},\ldots,d_{p})$, arising from the posterior $\pi(D_{-k}|{\bf X})$ for large $B$. Further $\tilde{s}_k\to \omega^{-1}_{0,kk}$, similarly as in the proof of Lemma 2 in Bondell et al. (2012). The posterior samples from $\pi(d_j|{\bf X}),j=1,\ldots,p,$ are $o_P(n)$, and hence condition (B3) is satisfied in a probabilistic sense, which is enough for Lemmas 1-4 in Bondell et al. (2012) to hold. As in Bondell et al. (2012), establishing Lemmas 1-4 leads to the proof of Theorem 3 for the conditional regression $X_{ck}|{\bf X}_{-k}, k=1,\ldots,p$. Now note that in our case, $A_n$ in Theorem 3 is equivalent to $\hat{\mbox{ne}}_{k,n}$ for the conditional regression $X_{ck}|{\bf X}_{-k}, k=1,\ldots,p$, which implies $P(\hat{\mbox{ne}}_{k,n} = \mbox{ne}_{0k}| {\bf X}_{-k}) \to 1$ as $n\to \infty$ almost surely with respect to ${\bf X}_{-k}$. Using dominated convergence Theorem, $P(\hat{\mbox{ne}}_{nk} = \mbox{ne}_{0k}) = E\left[P(\hat{\mbox{ne}}_{nk} = \mbox{ne}_{0k}| {\bf X}_{-k}) \right] \to 1$, so Theorem 2 is proved.

\vskip 12pt

{\noindent{\bf Proof of Corollary 1:}} Let $\mbox{ADJ}_0$ denote the adjacency matrix corresponding to the edge set $E_0$. Now note that $|\hat{\Omega}_{\hat{E}_n}- \Omega_{E_0}|\le |\hat{\Omega}_{\hat{E}_n} - \hat{\Omega}_{E_0}| + |\hat{\Omega}_{E_0} - \Omega_{E_0}|$, where $\hat{\Omega}_{E_0} = \hat{\Omega} \otimes ADJ_0$, with $ \hat{\Omega}$ denoting the posterior mean of $\Omega$. Since $d_k=o_P(n)$ as argued in the previous proof (k=1,\ldots,$p$), $\hat{\Omega}\approx (\frac{1}{n}({\bf X}^T{\bf X}))^{-1}\stackrel{a.s.}{\rightarrow} \Omega_{E_0}$, so that $\hat{\Omega}_{E_0}\stackrel{a.s.}{\rightarrow}  \Omega_{E_0}$ as $n\to \infty$. 
Further from the first part of Theorem 2, $P(E_n=E_0)\to 1$ as $n\to \infty$, which implies $P(\hat{\Omega}_{\hat{E}_n} = \hat{\Omega}_{E_0})\to 1$ as $n\to \infty$. The rest follows. \\

\singlespace

\pagenumbering{gobble}
\begin{small}
{\noindent\textbf{Table 1.1}: Area under ROC curve for true edge set ES005 (ROC) and true edge set ES1 (ROC1), with standard errors in parenthesis, under Case I. Results for $(n,p)=(300,100),(400,200),(500,100),(500,200)$, based on 50 replicates.}
\begin{center}
\begin{tabular}{|l | l l l l l l l l |}
\hline 
$(n,p)$ &\multicolumn{2}{c}{300,100} &\multicolumn{2}{c}{400,200}  &\multicolumn{2}{c}{500,100} &\multicolumn{2}{c}{500,200} \\
 \hline
                            &ROC  &ROC1          &ROC   &ROC1          &ROC &ROC1          &ROC &ROC1        \\ [10pt]   
RIW   $\times 10^{-2}$      &63(0.94) &98(0.39)  &67(0.70)  &97(0.22)  &66(0.99) &99(0.11) &67(0.6)  &99(0.2)  \\
GLASSO$\times 10^{-2}$      &62(0.95) &99(0.03)  &63(0.70)  &99(0.20)  &64(0.92) &99(0.01) &64(0.6)  &100(0.16) \\
IW    $\times 10^{-2}$      &58(0.79) &97(0.2)  &61(0.44)  &96(0.15)  &58(0.81) &99(0.18) &60(0.73) &99(0.13) \\
BGLA  $\times 10^{-2}$      &39(0.77) &68(1.2)   &37(0.58)  &66(0.19)  &46(0.88) &73(0.80) &39(0.66) &68(0.19)     \\
BGAD  $\times 10^{-2}$      &38(0.81) &74(0.6)   &36(0.61)  &72(0.23)  &42(0.91) &77(0.62) &39(0.69) &73 (0.21)     \\ 
HIW   $\times 10^{-2}$      &53(2) &81(6)   &57(3)  &91(3)  &51(3) &85(7) &58(2) &92(3)            \\ [10pt]
\hline
\end{tabular}
\end{center}
\end{small}

\begin{small}
{\noindent\textbf{Table 1.2}: Area under ROC curve for true edge set ES005 (ROC) and true edge set ES1 (ROC1), with standard errors in parenthesis, under Case II, for $n=200,300$ and $p=49$. Results based on 50 replicates.}
\begin{center}
\begin{tabular}{|l | l l l l|}
\hline 
n &\multicolumn{2}{c}{200} &\multicolumn{2}{c}{300}    \\
 \hline
                            &ROC  &ROC1          &ROC   &ROC1            \\ [10pt]   
RIW     $\times 10^{-2}$    &59(1) &87(0.78)       &64(1)  &88(0.8)             \\
GLASSO  $\times 10^{-2}$    &46(2) &75 (0.6)       &49(1)  &76(0.7)              \\
IW      $\times 10^{-2}$    &53(1) &75(1)          &57(2)  &82(2)               \\
BGLA    $\times 10^{-2}$    &32(1) &43(1)          &38(1)  &48(1)                \\
BGAD    $\times 10^{-2}$    &38(1) &60(1)          &43(2)  &67(1)                 \\  
HIW     $\times 10^{-2}$    &55(2) &82(2)          &57(1)  &82(1)         \\[10pt]
\hline
\end{tabular}
\end{center}
\end{small}

\begin{small}
{\noindent\textbf{Table 2.1}: Estimated specificity and sensitivity percentages for the best graph under Case I. Results for $(n,p)=(300,100),(400,200),(500,100),(500,200)$. SP, SE, and SP1, SE1, denote specificity and sensitivity under true edge sets ES005 and ES1 respectively. Results based on 50 replicates.}
\begin{center}
\begin{tabular}{|l | l l l l l l l l l l l l l l l l|}
\hline 
 $(n,p)$ &\multicolumn{4}{c}{300,100} &\multicolumn{4}{c}{400,200}  &\multicolumn{4}{c}{500,100} &\multicolumn{4}{c}{500,200} \\
 \hline
          &SP &SE  &SP1 &SE1      &SP &SE     &SP1 &SE1        &SP &SE   &SP1 &SE1     &SP &SE   &SP1  &SE1   \\      
RIW       &95 &25  &92  &100      &98 &18     &99   &99        &92 &33   &90   &100     &98 &24   &97   &100      \\
GLASSO    &99 &21  &96   &99      &99 &21     &99   &100       &97 &31   &95   &100     &99 &25   &97   &100      \\
IW        &97 &14  &95   &91      &94 &19     &97   &99        &98 &14   &93   &99      &98 &11   &98   &99       \\
BGLA      &80 &34  &88   &58      &87 &28     &83  &95        &74 &47   &81   &99      &85 &33   &84  &99        \\
BGAD      &94 &19  &95   &91      &95 &19     &94  &99        &95 &23   &93   &100     &95 &21   &95  &100         \\
HIW       &100 &07  &99   &84      &100 &06   &100  &70     &99 &08   &99   &89   &99 &07   &99   & 64         \\
\hline
\end{tabular}
\end{center}
\end{small}

\begin{small}
{\noindent\textbf{Table 2.2}: Estimated specificity and sensitivity percentage for the optimal graph under all approaches under Case II, for $n=200,300$ and $p=49$. SP, SE, and SP1, SE1, denote specificity and sensitivity under true edge sets ES005 and ES1 respectively. Results based on 50 replicates.}
\begin{center}
\begin{tabular}{|l | l l l l l l l l|}
 \hline 
 n &\multicolumn{4}{c}{200} &\multicolumn{4}{c}{300}  \\
 \hline
          &SP &SE  &SP1 &SE1 &SP &SE &SP1 &SE1           \\      
RIW       &85 &30  &79 &75   &84 &35 &77  &84             \\
GLASSO    &92 &04  &93 &09   &93 &03 &95  &09              \\
IW        &87 &19  &83 &36   &88 &25 &83  &56                 \\
BGLA      &57 &34  &62 &47   &45 &49 &48  &74               \\
BGAD      &94 &16  &91 &56   &94 &21 &88  &68                \\
HIW       &99 &03  &99 &25   &100 &04 &99  &36            \\
\hline
\end{tabular}
\end{center}
\end{small}

\begin{small}
{\noindent\textbf{Table 3}: Area under ROC curve (with standard errors in parenthesis) and sensitivity and specificity percentages under the optimal graph for RIW and GLASSO under Case I, for $(n,p)=(700,500)$. Results based on 50 replicates.}
\begin{center}
\begin{tabular}{|l | l l l l l l |}
\hline
            &ROC                  &ROC1                   &SP   &SE       &SP1 &SE1    \\ [10pt] 
RIW         &70$\times 10^{-2}$   &100$\times 10^{-2}$    &99   &18       &99  &96   \\
GLASSO      &67$\times 10^{-2}$   &100$\times 10^{-2}$    &99   &25       &99  &99   \\            
\hline
\end{tabular}
\end{center}
\end{small}

\begin{small}
{\noindent\textbf{Table 4}: Computation times (in cpu secs) under Case I, for RIW, BGLA, BGAD approaches (10000 iterations) and HIW (100000 iterations). Results based on 50 replicates.}
\begin{center}
\begin{tabular}{|l| l l l l|}
 \hline   
   $(n,p)$           &(300,100) &(400,200)  &(500,100)     &(500,200)   \\
  RIW $\times10^4$   &0.008    &0.24          &0.01      &0.33                  \\
  BGLA $\times10^4$  &4.14     &$>10$         &5.66      &  $>10$                     \\
  BGAD $\times10^4$  &6.12     &$>10$         &6.57      & $>10$                 \\
  HIW$\times10^4$    &5.08     &$>10$         &6.07       &$>10$ \\
\hline
\end{tabular}
\end{center}
\end{small}

\begin{small}
{\noindent\textbf{Table 5.1}: Hub nodes ($>8$ neighbors) with number of neighbors for mRNA example.}
\begin{center}
\begin{tabular}{|l| l l l l l l |}
 \hline 
 Name          &EGFR &RAF1 &NF1 &SPRY2 &CDKN2A &PIK3C2G \\
 \# Neighbors  &9    &11   &9   &9     &9      &14      \\
 \hline
 \end{tabular}
\end{center}
\end{small}

\begin{small}
{\noindent\textbf{Table 5.2}: Hub nodes ($>10$ neighbors) with number of neighbors for miRNA example.}
\begin{center}
\begin{tabular}{|l| l l l l l|}
 \hline 
 Name       & ebv-mir-bart7 &hsa-mir-106a &hsa-mir-142-3p &hsa-mir-17-5p  &hsa-mir-let7 \\
 \# Neighbors  &11    &26   &13   &14   &26     \\
 \hline
 Name &hsa-mir-181c &hsa-mir-184 &hsa-mir-19a &hsa-mir-20a & \\
 \# Neighbors  &12      &15  &13   &11   & \\
 \hline
 \end{tabular}
\end{center}
\end{small}

\begin{figure}
		\mbox{\includegraphics[height=3.5in, width=0.5\textwidth]{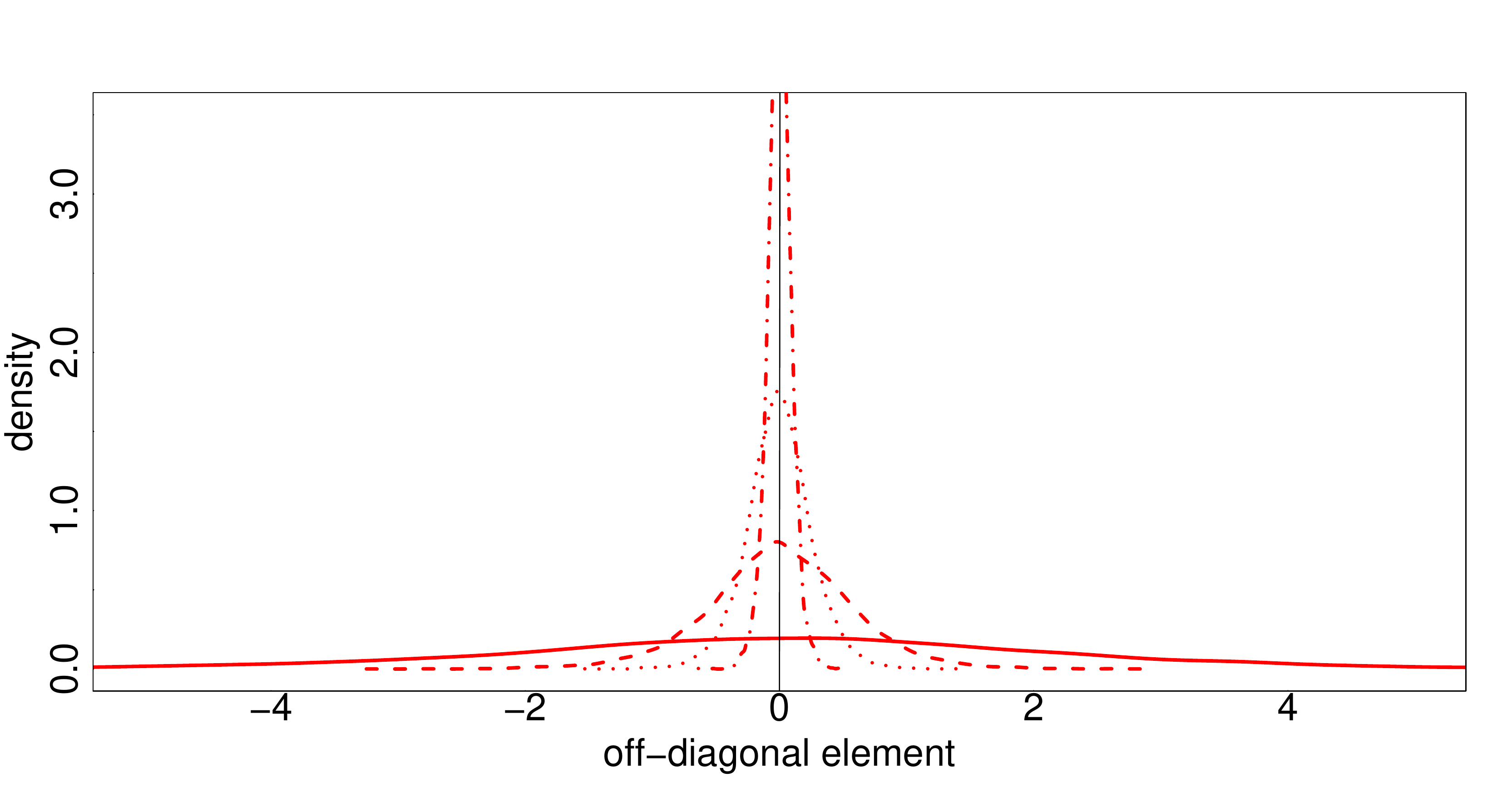} \quad \includegraphics[height=3.5in, width=0.5\textwidth]{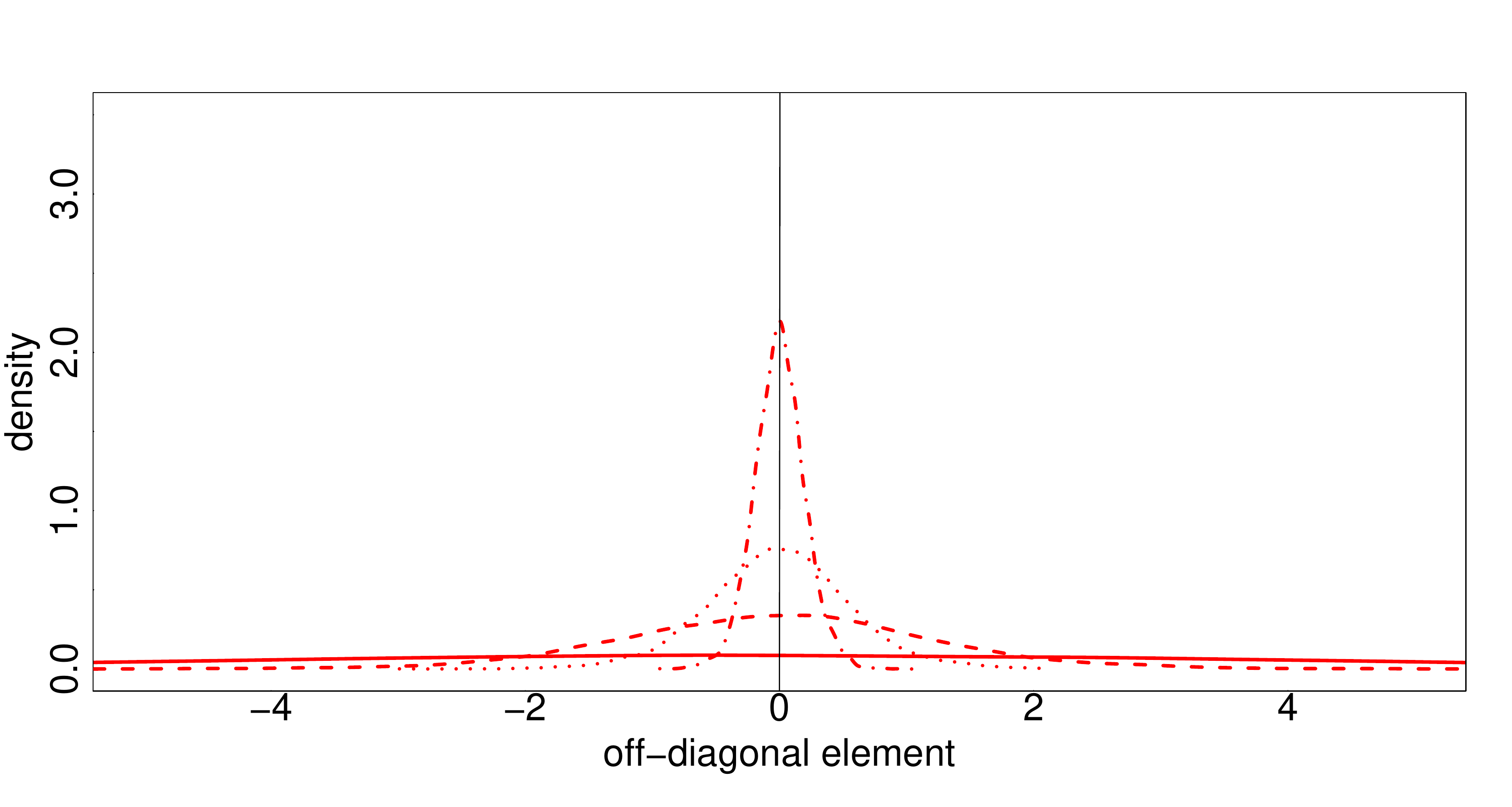}} 
			\caption{\small{Prior realizations for precision off-diagonals under prior formulation (\ref{eq:modbase1}) for $\lambda=5$ (solid), $\lambda=10$ (dashed), $\lambda=15$ (dotted), and $\lambda=25$ (dots and dashes). Left plot has $p=10$, right plot has $p=20$, while $b=p$ in all plots.}}
\end{figure}

\begin{figure}
		\mbox{\includegraphics[height=7.5in, width=1\textwidth]{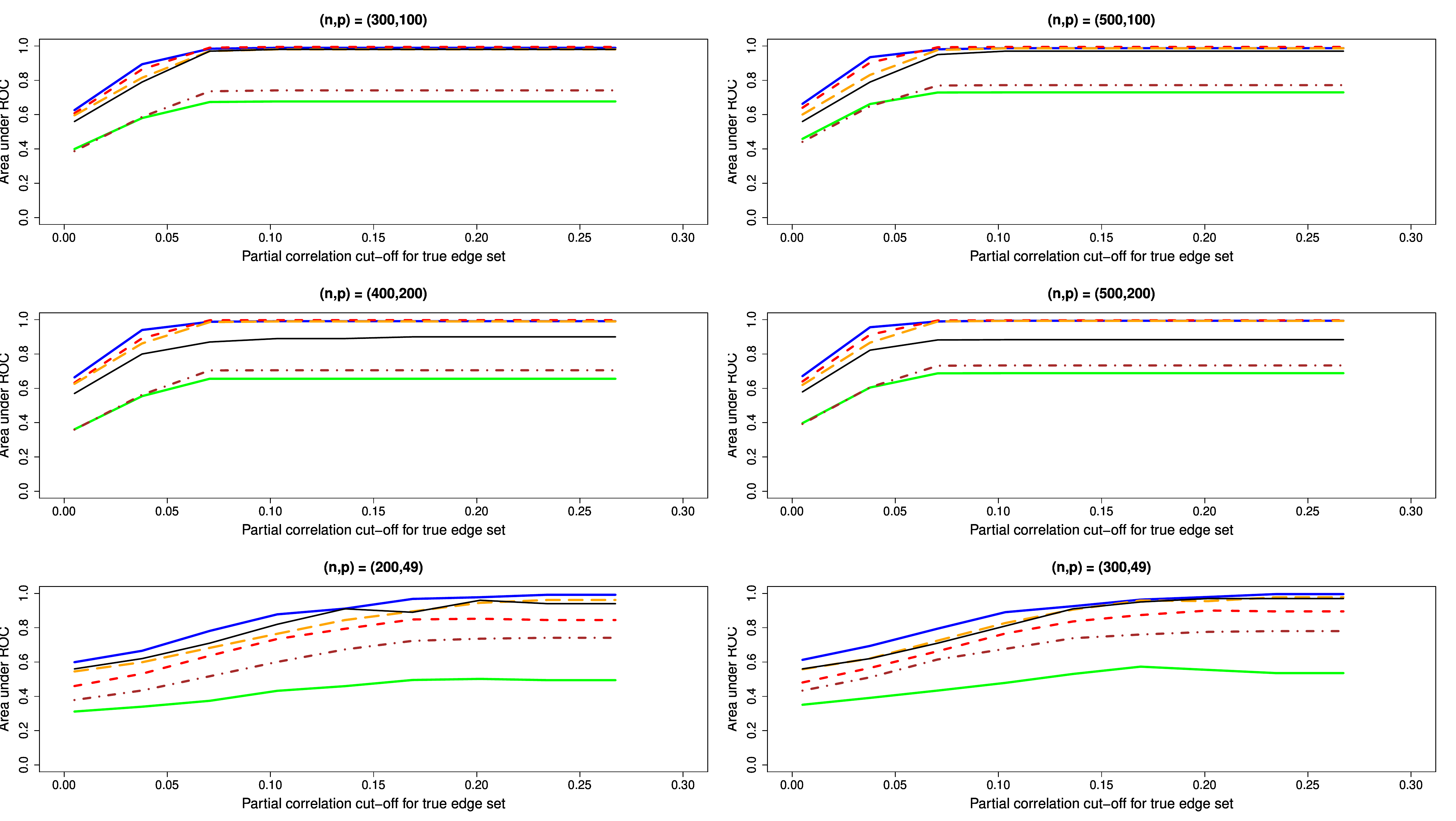} }
			\caption{\small{Area under ROC curves across different absolute partial correlation thresholds, for true edge set ES1. Thick solid line is RIW, short dashes is Glasso, dotted line is BGLA, dots \& dashes are BGAD, long dashes is IW, thin solid line is HIW. Blue solid line is RIW, red dashes is Glasso, green dotted line is BGLA, brown dots \& dashes are BGAD, orange long dashes is IW, thin solid black line is HIW.}}
\end{figure} 

\begin{figure}
		\mbox{\includegraphics[height=7.5in, width=1\textwidth]{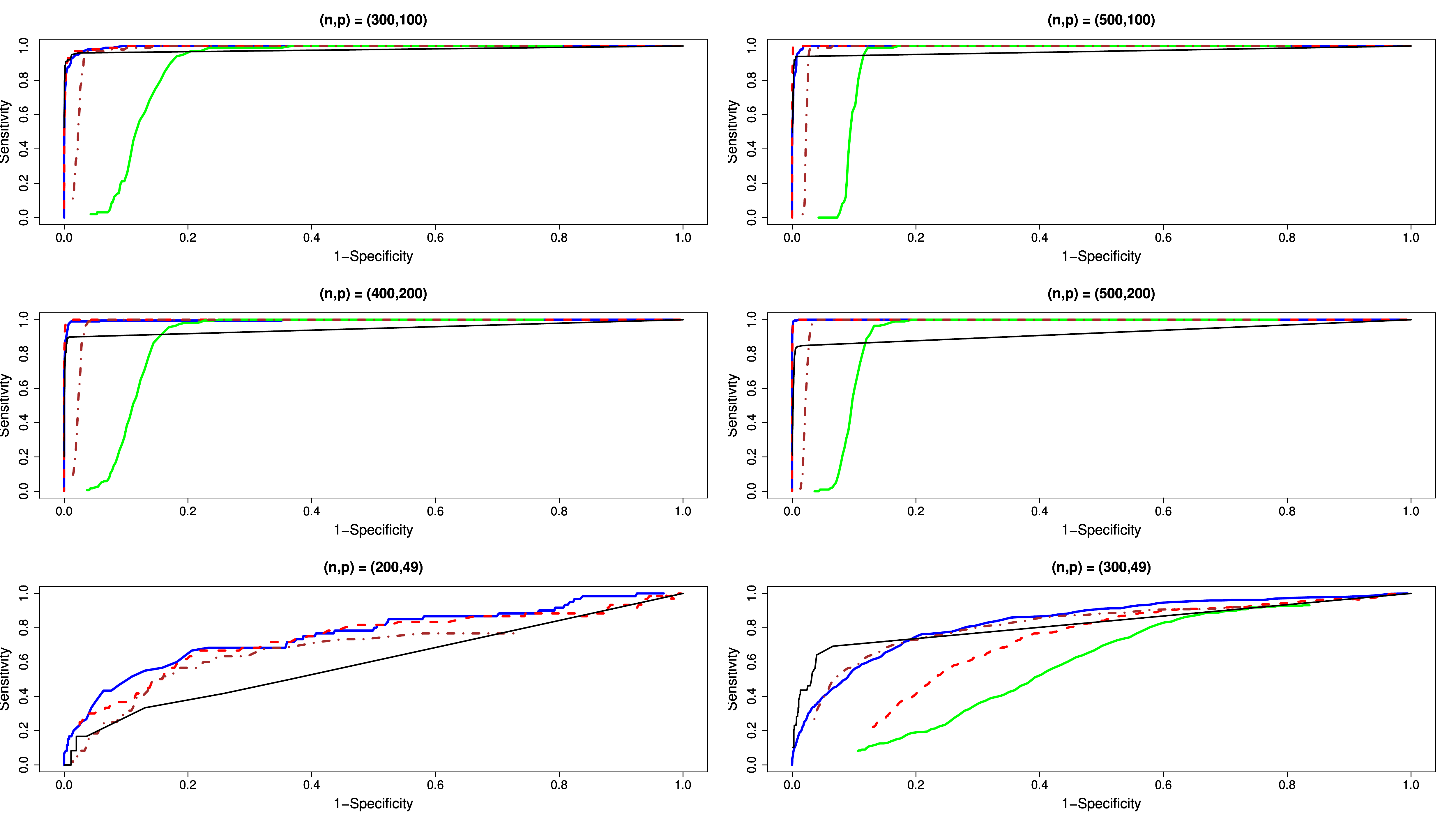} }
			\caption{\small{ROC curves for true edge set ES1. Top four panels are for Case I and bottom two panels are for Case II. Blue solid line is RIW, red dashes is Glasso, green dotted line is BGLA, brown dots \& dashes are BGAD, orange long dashes is IW, thin solid black line is HIW.}}
\end{figure}  

\begin{figure}
		\mbox{\includegraphics[height=7in, width=1\textwidth]{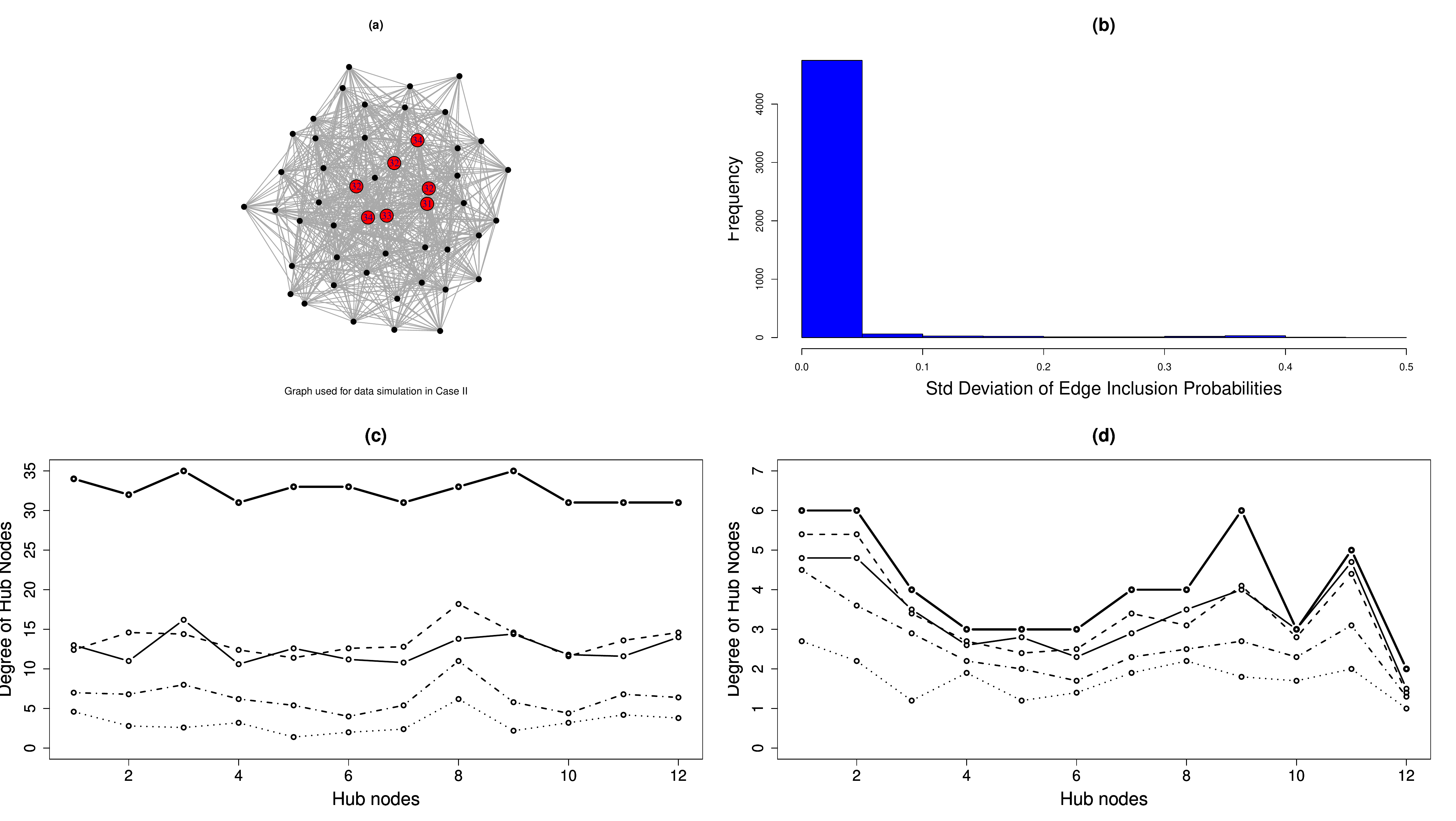} }  
			\caption{\small{Panel (a): Graph used for Case II, with hub nodes made prominent and labeled using the degree. Panel (b): Standard deviation of edge inclusion probabilities for HIW for Case II when n=100; Panel (c): Estimated degree for hub nodes for Case II when true edge set is ES005; Panel (d): Estimated degree for hub nodes for Case II when true edge set is ES1. For Panels (c)-(d), thick solid line is true degree, solid line is degree under RIW, dotted line is GLASSO, dashed line is BGLA, and dots and dashes is BGAD. }}
\end{figure}   

\begin{figure}
		\mbox{\includegraphics[angle = 90, height=7in, width=1\textwidth]{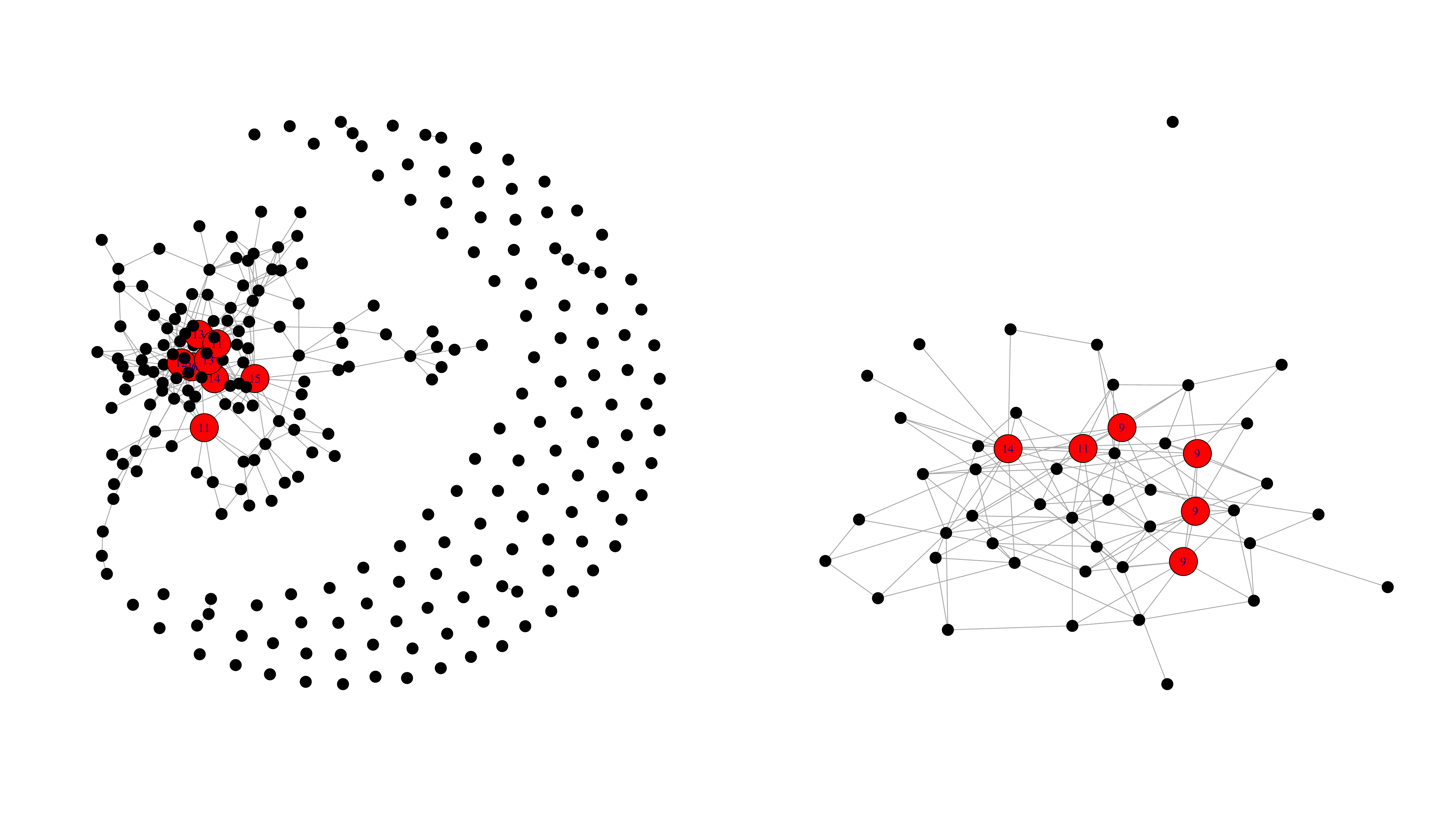} }
			\caption{\small{Top: Estimated graph for mRNA expression data; Bottom: Estimated graph for miRNA expression data. Large nodes indicate hub nodes, with the number of neighbors labeled inside the hub nodes.}}
\end{figure}

\end{document}